\documentclass[twocolumn]{aastex63}

\usepackage{gensymb}
\usepackage{bm}

\accepted{for ApJ}

\shorttitle{Habitable World Polarization}
\shortauthors{Gordon et al. 2022}
\graphicspath{{./}{figures/}}

\begin{document}

\title{Polarized Signatures of a Habitable World: \\ Comparing Models of an Exoplanet Earth with Visible and Near-infrared Earthshine Spectra}

\correspondingauthor{Kenneth E. Gordon}
\email{ke14gordon@knights.ucf.edu}

\author[0000-0002-4258-6703]{Kenneth E. Gordon}
\affiliation{Planetary Sciences Group, Department of Physics, University of Central Florida, 4111 Libra Drive, Orlando, FL 32816, USA}

\author[0000-0001-7356-6652]{Theodora Karalidi}
\affiliation{Planetary Sciences Group, Department of Physics, University of Central Florida, 4111 Libra Drive, Orlando, FL 32816, USA}

\author[0000-0002-4420-0560]{Kimberly M. Bott}
\affiliation{Department of Earth and Planetary Sciences, University of California, Riverside, Riverside, CA 92521, USA}
\affiliation{NASA Nexus for Exoplanet System Science, Virtual Planetary Laboratory Team, Box 351580, University of Washington, Seattle, WA 98195, USA}

\author[0000-0003-2446-8882]{Paulo A. Miles-Páez}
\altaffiliation{ESO Fellow}
\affiliation{European Southern Observatory, Karl-Schwarzschild-Straße 2, D-85748, Garching bei München, Germany}

\author[0000-0001-5762-9385]{Willeke Mulder}
\affiliation{Leiden Observatory, Leiden University, P.O. Box 9513, 2300 RA Leiden, The Netherlands}

\author[0000-0003-3697-2971]{Daphne M. Stam}
\affiliation{Faculty of Aerospace Engineering, Astrodynamics and Space Missions, Delft University of Technology, Kluyverweg 1, 2629 HS, Delft, The Netherlands}



\begin{abstract}
 
In the JWST, Extremely Large Telescopes, and LUVOIR era, we expect to characterize a number of potentially habitable Earth-like exoplanets. However, the characterization of these worlds depends crucially on the accuracy of theoretical models. Validating these models against observations of planets with known properties will be key for the future characterization of terrestrial exoplanets. Due to its sensitivity to the micro- and macro-physical properties of an atmosphere, polarimetry will be an important tool that, in tandem with traditional flux-only observations, will enhance the capabilities of characterizing Earth-like planets. In this paper we benchmark two different polarization-enabled radiative-transfer codes against each other and against unique linear spectropolarimetric observations of the earthshine that cover wavelengths from $\sim$0.4 to $\sim$2.3~$\mu$m. We find that while the results from the two codes generally agree with each other, there is a phase dependency between the compared models. Additionally, with our current assumptions, the models from both codes underestimate the level of polarization of the earthshine. We also report an interesting discrepancy between our models and the observed 1.27~$\mu$m $O_2$ feature in the earthshine, and provide an analysis of potential methods for matching this feature. Our results suggest that only having access to the 1.27~$\mu$m $O_2$ feature coupled with a lack of observations of the $O_2$ A and B bands could result in a mischaracterization of an Earth-like atmosphere. Providing these assessments is vital to aid the community in the search for life beyond the solar system.

\end{abstract}



\section{Introduction} \label{sec:intro}


Within the last decade, exoplanet missions including Kepler and the Transiting Exoplanet Survey Satellite have been aimed at determining $\eta_\earth$, the occurrence rate of habitable-zone (HZ) rocky exoplanets around solar-type stars. Recently, \citet[][]{bryson2020} combined statistical data from the Kepler Data Release 25 planet candidate catalog along with stellar properties measured by the Gaia mission and found that $\eta_\earth$ for the conservative HZ around main-sequence dwarf stars ranges from $0.37^{+0.48}_{-0.21}$ to $0.60^{+0.90}_{-0.36}$ planets per star, while for the optimistic HZ, $\eta_\earth$ ranges from $0.58^{+0.73}_{-0.33}$ to $0.88^{+1.28}_{-0.51}$ planets per star. Terrestrial exoplanets are therefore expected to be frequent, with more than 180 rocky planets confirmed thus far (as per the NASA Exoplanet Archive). These rocky exoplanets provide for intriguing studies as astronomers strive to discover life on distant worlds. However, to know whether a rocky exoplanet is habitable, we need to be able to characterize its atmosphere and surface.

Currently, transit spectroscopy and the direct detection of planetary flux are the most widely used techniques for performing these characterizations. While proven successful for the gaseous giant exoplanets at very short \citep[transiting; e.g.,][]{line2013,sing2016} or very wide {\citep[imaging; e.g.,][]{marois2008, desgrange2022}} orbital distances, Earth-sized exoplanets around Sun-like stars are more challenging and introduce a number of issues. In transit, the smaller planetary sizes and wider orbital distances of habitable terrestrial planets compared to gaseous giant planets mean that they only block a minuscule fraction of their host star upon transiting \citep[e.g.,][]{kaltenegger2009, palle2011}. Additionally, the longer orbital periods of terrestrial planets around solar type stars require more time for follow-up studies to confirm any possible transits, and their thinner atmospheres complicate spectroscopic characterization \citep[e.g.,][]{betremieux2014, misra2014}. For direct detection, the main issue lies in the minuscule amount of emitted and/or reflected flux from the small planet compared to its host star \citep[e.g.,][]{traub2010, seager2014}. Therefore, full characterization of the atmospheres and surfaces of terrestrial exoplanets in the HZs of solar-type stars remains elusive using these traditional methods.


The recent launch of the JWST will help to alleviate some of these issues. JWST will allow for numerous follow-up characterizations of transiting planets across a range of masses and atmospheres \citep[e.g.][]{greene2016}. However, debate still exists on the feasibility of JWST to effectively characterize the atmospheres of potentially habitable Earth-like planets orbiting cool dwarf stars, mainly due to the necessity for long observational timeframes \citep[see e.g.][and references therein]{robinson2018, wunderlich2019}. For example, \citet[][]{barstow2016} analyzed the three potentially habitable Earth-like planets around TRAPPIST-1: planets 1b, 1c, and 1d. Using JWST's Near-InfraRed Spectrograph (NIRSpec) and Mid-InfraRed Instrument (MIRI), they found that $\gtrsim$60 transits for 1b and $\gtrsim$30 transits for 1c and 1d would be required to detect present-day Earth levels of ozone (O$_3$) on these planets.

Recently, \citet[][]{gialluca2021} showed that the detectability of biosignature gases at specific wavelength bins (i.e., only considering the central wavelength of strong absorption features) will be very challenging for JWST. However, integrating the spectral impact of a gas across the entire wavelength range of an instrument improves the observational time constraints and could allow for detections of $CH_4$, $CO_2$, $O_2$, and $H_2O$ with a few tens of transits for a cloud-free Earth analog orbiting cool dwarf stars up to 15 pc away. \citet[][]{gialluca2021} stressed, though, that their results are highly dependent on the assumed planetary atmospheric chemistry and host star environment of the models, so more in-depth analyses are required. Additionally, \citet[][]{gialluca2021} did not include clouds in their models, which can severely limit the detectability of key biosignatures such as $H_2O$ \citep[e.g.,][]{suissa2020}. Therefore, while JWST will provide some assistance, a number of degeneracies will still exist in the characterizations of habitable worlds by JWST, such as differentiating between the optical thicknesses and particle-size distributions of clouds.


Polarimetry is a powerful technique that has the ability to break these degeneracies, as it assesses physical aspects of light not measured in nonpolarimetric photometry or spectroscopy. Polarimetry measures light as a vector (by the orientation of the electric field oscillations) rather than only a scalar intensity, thus making it extremely sensitive to the physical and microphysical properties of an atmosphere \citep[e.g.,][]{hansenhovenier1974, hansentravis1974}. Due to the vector nature of polarimetry, it is also sensitive to the location of specific features on the disk of an object \citep[see e.g.,][]{karalidi2013flux, stolker2017}. Polarimetry has helped to characterize bodies in the Solar System, including the  clouds of Venus \citep{hansenhovenier1974} and the gas giants \citep{schmid2011} as well as the differing icy conditions of the Galilean Moons \citep[]{dollfus1975, rosenbush2002}. Finally, polarimetry is especially useful in identifying whether reflection is coming from an absorptive surface or a cloud deck, making it highly applicable to studies of habitable worlds \citep[e.g.,][]{fauchez2017}.


Various groups have modeled the flux \citep[e.g.,][]{lincowski2018, meadows2018proxima, lustig2019, leung2020} and linear polarimetric signals \citep[e.g.,][]{stam2008, karalidi2011, karalidi2012rainbow, karalidistam2012, bailey2018, treesstam2019, groot2020, trees2022} of Earth-like planets as functions of orbital phase and wavelength. Models, however, need to be validated against observational data. To date, the Earth is the only known and observed habitable ``Earth-like'' planet, thus serving as a benchmark to infer the biosignatures of life as we know it today.

The best current method of studying the Earth-as-an-exoplanet is through observations of the earthshine: sunlight scattered or reflected by the dayside of the Earth and reflected back to the planet by the nightside of the Moon, where it can then be measured by ground-based facilities. Studies of the optical and near-infrared (NIR) earthshine flux spectra \citep[e.g.,][]{woolf2002, turnbull2006, palle2009} reveal diagnostic biosignatures of the Earth, including the vegetation red edge (VRE), the ocean glint, and spectral features of atmospheric $O_{2}$ and $H_{2}O$. Studies on the spectropolarization of the earthshine \citep{dollfus1957, sterzik2012, sterzik2019, sterzik2020, bazzon2013, takahashi2013, takahashi2021, milespaez2014} detected the same biosignatures, and showed the sensitivity of polarization to features such as water clouds, varying surfaces, and ocean glint.



With the increased interest and expected near-future surplus of available data for terrestrial exoplanets, efforts have been made to benchmark and validate different codes focused on modeling these planets \citep[e.g.,][]{fauchez2021, paradise2022}. Here, we focus on the polarization of the reflected light from terrestrial planets and provide the first benchmarking of two independent polarized radiative-transfer codes against each other and against observations. In particular, we present the first comparisons of theoretical models from the Doubling Adding Program (DAP) code \citep{deHaan1987, stam2008} and the Versatile Software for Transfer of Atmospheric Radiation \citep[VSTAR;][]{spurr2006, bailey2018} against each other for a range of cloud and surface properties, and then compare these models against the visible to near-infrared (VNIR) observed earthshine spectrum presented in \citet{milespaez2014}. This spectrum provided the first extension of the wavelength range for linear spectropolarimetry of earthshine into the NIR regime, reporting the first detection of NIR $H_{2}O$ around 0.93 and 1.12~$\mu$m and NIR $O_{2}$ around 1.27~$\mu$m in polarized earthshine data. To create our model exoplanet-Earth we utilized cloud and land cover properties derived from observations by the Moderate Resolution Imaging Spectroradiometer (MODIS) instrument on-board NASA's Terra and Aqua satellites \citep{king2004}, and corrected the observations for lunar depolarization.


The outline of this paper is as follows. Section \ref{sec:data} presents the observational data as well as the satellite data used to produce our models. In Section \ref{sec:codes}, we provide short descriptions of the two different codes that we compare in this paper. Then, in Section \ref{sec: clouds} we highlight the importance of clouds in modeling terrestrial exoplanets and the effects that changes in cloud parameters have on the resulting polarization signals. Section \ref{sec: codecompare} compares the resulting degree of polarization from the two codes against each other and the earthshine observations. In Section \ref{sec: o2feat} we provide further analysis on the 1.27~$\mu$m $O_2$ band in the earthshine spectrum. Finally, in Section \ref{sec:sum}, we summarize our results and describe possible improvements to the different numerical simulations, as well implications of our models on future missions aimed at enhancing our understanding of Earth-like exoplanets.

\section{The Data} \label{sec:data}

\subsection{Earthshine Observations} \label{sec: observe}

\begin{figure}[ht!]
    \centering
    \includegraphics[width=\linewidth]{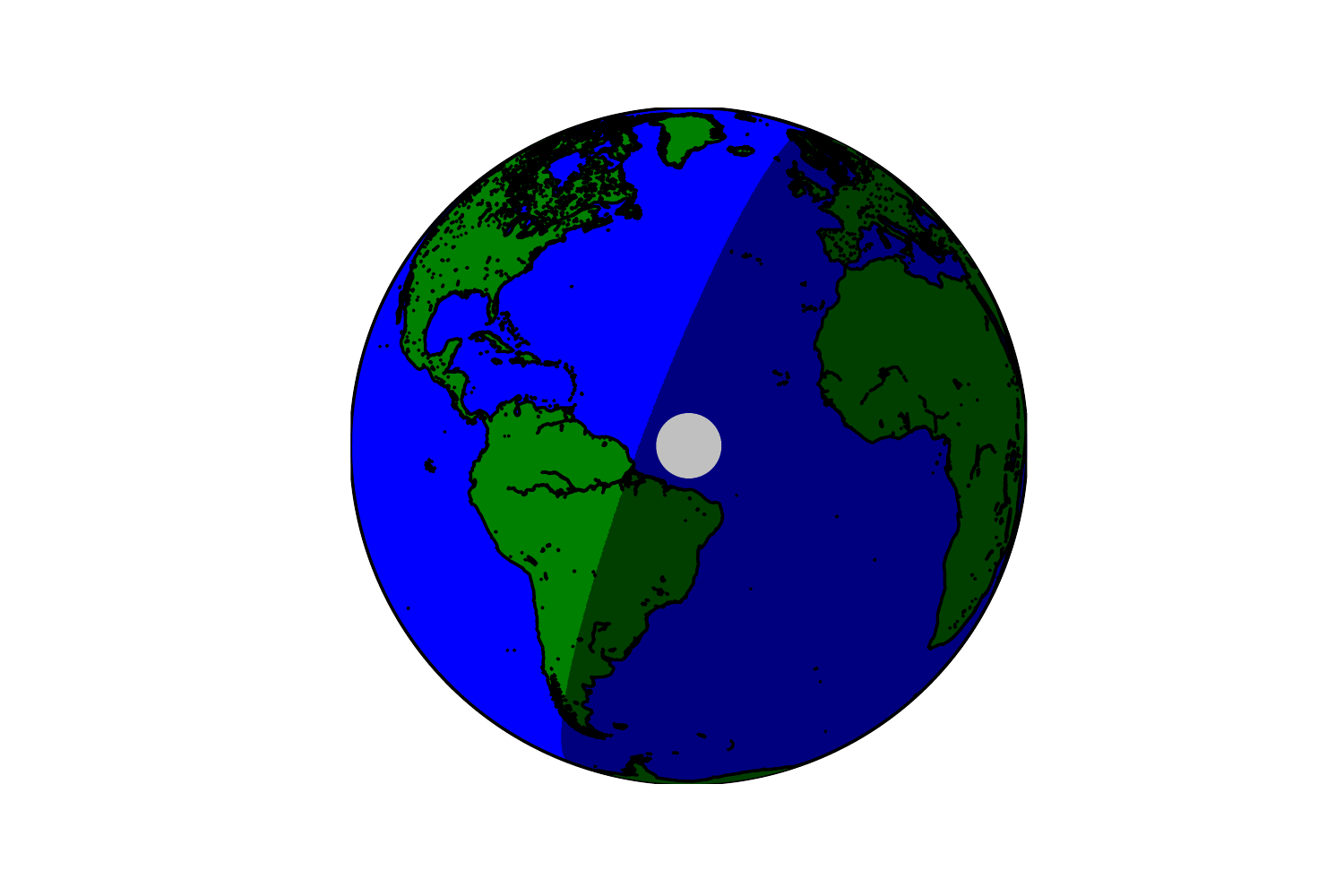}
    \centering
    \includegraphics[width=\linewidth]{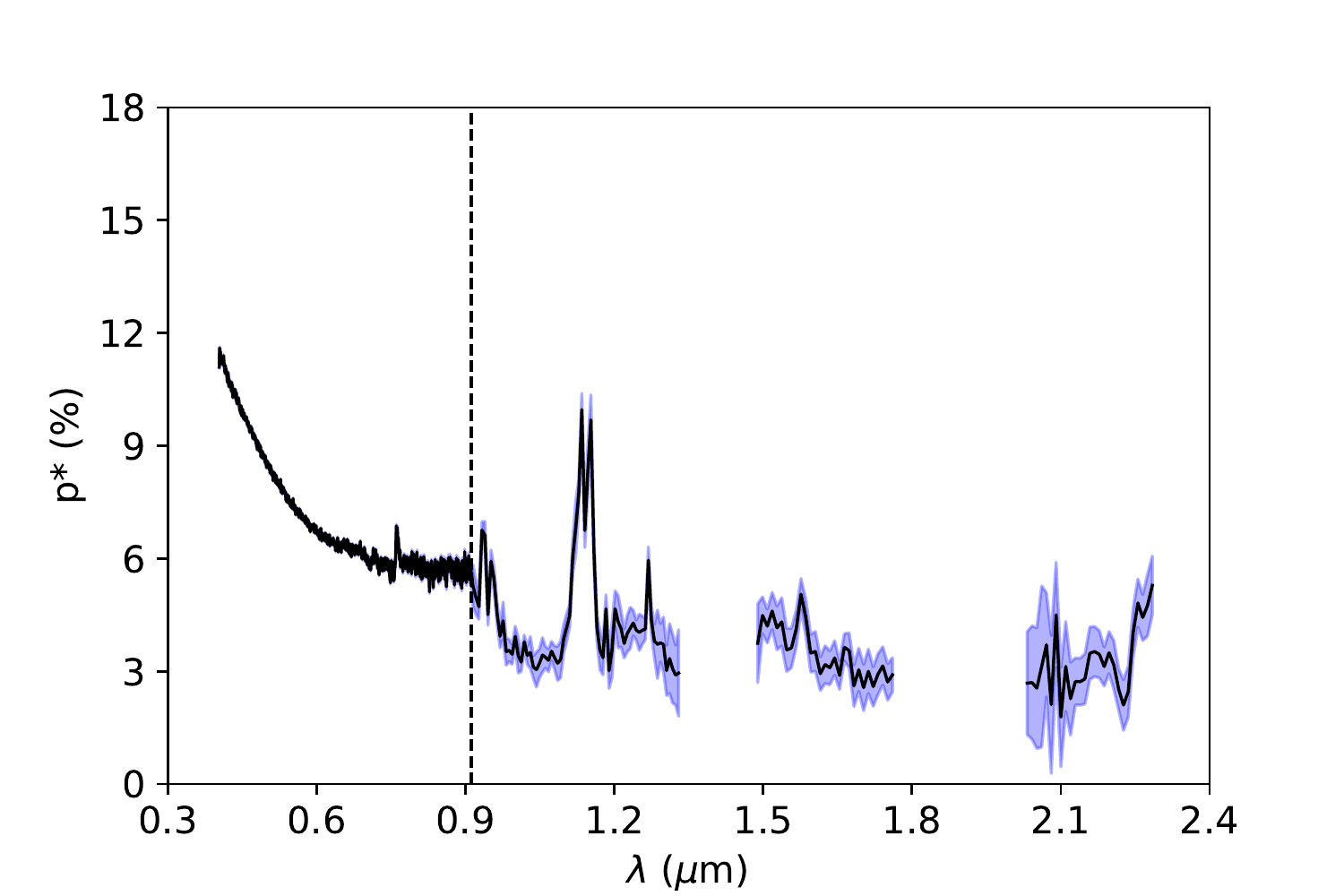}
    \caption{Top: the zenith location of the Moon on planet Earth at 21:20 UT on 18 May 2013: latitude $4\degree$ 36' North, longitude $40\degree$ 36' West. The lit portion of the globe is what was visible from the lunar surface. Bottom: the VNIR earthshine observations as published in \citet{milespaez2014}, showing the debiased degree of linear polarization $p^{*}$ as a function of wavelength $\lambda$. Wavelengths of strong telluric absorption have been removed. The vertical dashed line around 0.9~$\mu$m separates the ALFOSC and LIRIS data. The uncertainty per wavelength is given by the blue shaded region.}
    \label{fig:Moonloc}
\end{figure}

\citet{milespaez2014} provided the first linear spectropolarimetric measurements of the earthshine in the VNIR wavelengths. \citet{milespaez2014} utilized the Andalucía Faint Object Spectrograph and Camera (ALFOSC) instrument of the 2.56 m Nordic Optical Telescope (NOT) as well as the Long-slit Intermediate Resolution Infrared Spectrograph (LIRIS) of the 4.2 m William Herschel Telescope (WHT) to capture simultaneous optical (ALFOSC; 0.4 - 0.9~$\mu$m, spectral resolution of 2.51 nm) and NIR (LIRIS; 0.9 - 1.4~$\mu$m, spectral resolution of 1.83 nm; 1.4 - 2.4~$\mu$m, spectral resolution of 2.91 nm) linear spectropolarimetric observations of the earthshine, thereby extending the knowledge of linear polarimetric earthshine data into redder wavelengths than previous studies and allowing for the detection of important biosignature features not found at visible wavelengths. To increase the signal-to-noise ratio (S/N) of the observations in the NIR, they applied a 10 pixel binning to the spectrum in this wavelength region. The final spectrum therefore had a variable resolving power R ($= \lambda / \Delta\lambda$) of $\sim250$ at $\lambda = 0.65$~$\mu$m and $R \sim208$ at $\lambda = 1.27$~$\mu$m. For more information on the observations and calibrations of the data, we refer the reader to \citet{milespaez2014}.

\begin{figure}[]
    \centering
    \includegraphics[width=\linewidth]{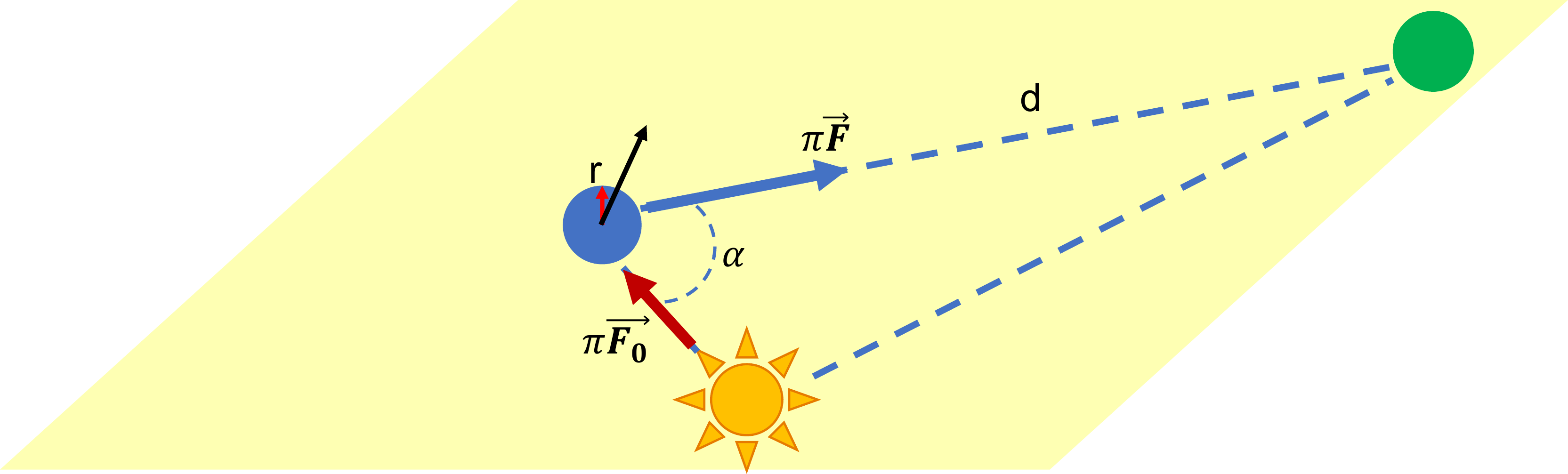}
    \newline \newline
    \centering
    \includegraphics[width=\linewidth]{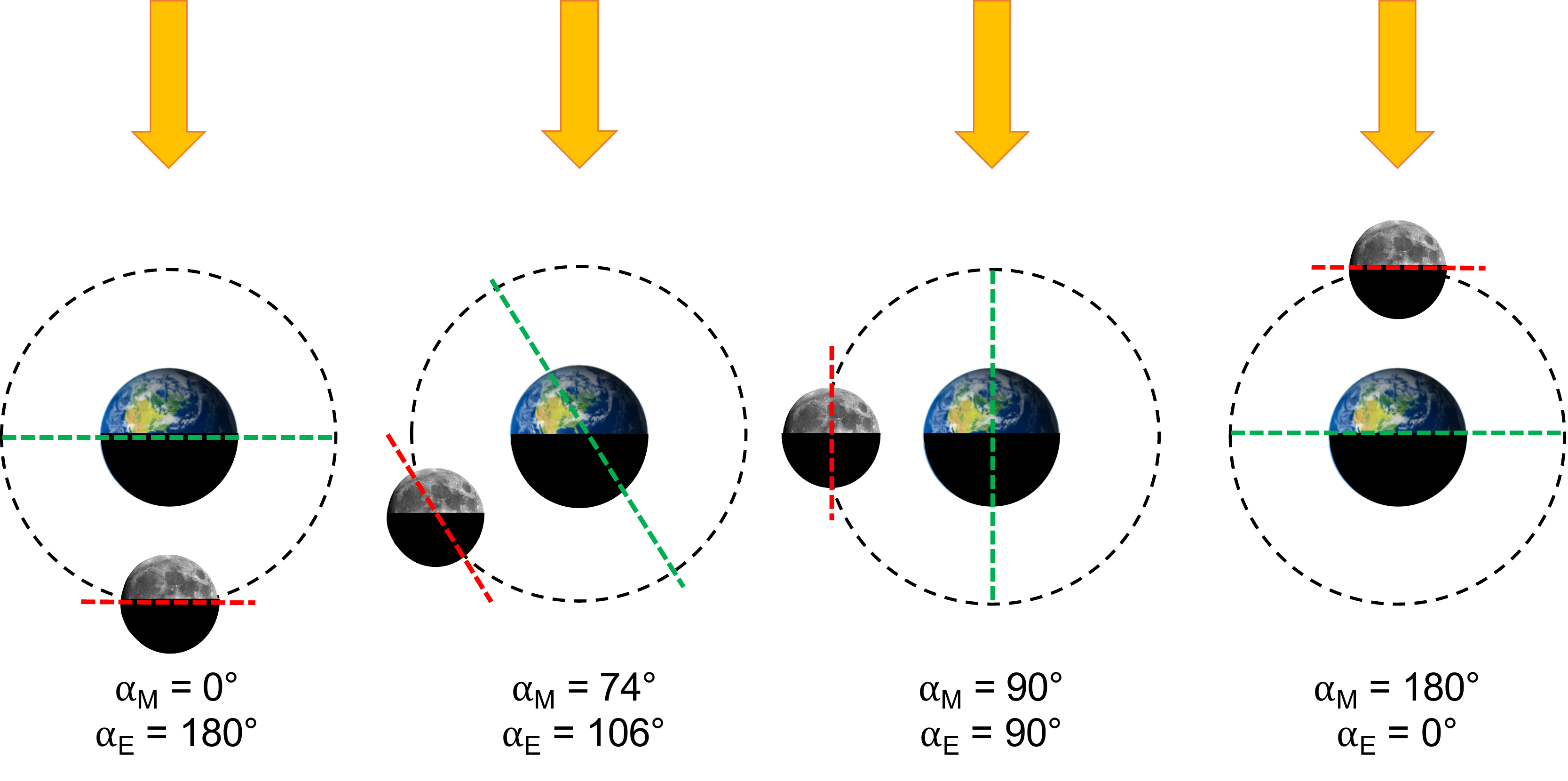}
    \caption{Top: the planetary scattering plane defines the plane through the centers of the star, the planet (blue circle), and the observer (green circle). Here, the observed planet is assumed to be spherical with radius $r$ and located at distance $d (\gg r)$ from the observer.
    The flux vector reflected by the planet towards the observer ($\pi\textbf{F}$) depends on the incident flux vector from the star ($\pi\textbf{F}_0$) and the planetary phase angle $\alpha$. The black arrow represents the tilt of the reference plane for the earthshine observations with respect to the planetary scattering plane.
    Bottom: schematic diagram of how the phases of the Earth mirror the phases of the Moon as the latter orbits the former. The yellow arrows indicate the incident sunlight. The red (green) lines define the extent of the lunar (Earth) surface that can be seen from the Earth (Moon). In the leftmost (rightmost) panel, the Moon disk as seen from the Earth is illuminated 100\% (0\%), corresponding to a lunar phase $\alpha_M$ = $0^{\circ}$ ($180^{\circ}$), while the Earth disk as seen from the Moon is illuminated 0\% (100\%), corresponding to an Earth phase $\alpha_E$ = $180^{\circ}$ ($0^{\circ}$). The in-between phases are complementary to each other.}
    \label{fig:plane}
\end{figure}

Figure~\ref{fig:Moonloc} shows the Earth as observed from the Moon during the \citet{milespaez2014} observations (top) and the observed polarimetric earthshine spectrum (bottom). 
The reference plane of the observations is tilted by $29^\circ$ from the planetary scattering plane, which is the plane through the centers of the star, planet, and observer (see top panel of Figure~\ref{fig:plane}). The earthshine observations were obtained on a night when the waxing Moon illuminated area was $59\%$ \citep[][]{milespaez2014}. Due to the geometry of the Earth-Moon system, Earth phases are exactly opposite of lunar phases (see bottom panel of Figure~\ref{fig:plane}). This means that the Earth would have been illuminated $41\%$ on this night (as seen from the Moon), which corresponds to a phase angle of $\alpha = 106^{\circ}$ during the observations.
During the observations  
the largest contributors to the earthshine were the western Atlantic Ocean and the Amazon rainforest. In the wavelength range covered by these observations a number of strong absorption features exist, including those of atmospheric $O_{2}$ and $H_{2}O$, which are important biomarkers. Additionally, the presence of the Amazon rainforest in the field-of-view leaves a mark on the observations, with a small yet detectable VRE near $\lambda = 0.7$~$\mu$m.

\subsection{Exoplanet-Earth Models} \label{sec: MODIS}

To model the horizontal inhomogeneity of the visible Earth disk, we divided our model planet into pixels of $2\degree \times 2\degree$ such that the local properties of the atmosphere and surface are plane-parallel and horizontally homogeneous. These pixels are large enough to be able to ignore effects from surrounding pixels (e.g., light that is reflected or scattered by clouds in one pixel towards the surface of an adjacent pixel). We then ran each pixel through a radiative-transfer code which produces the full spectropolarimetric signal of the pixel across all wavelengths $\lambda$ in the VNIR (0.4 - 2.3~$\mu$m) and all phase angles $\alpha$ from $0\degree - 180\degree$ in steps of 2\degree. After running every pixel in the illuminated section of the Earth disk, we used a weighted averaging to combine the signals from all pixels and generate the disk-integrated linear polarization spectrum for our model exoplanet-Earth.

To test that our results are independent of the code we use, we compared the output of two separate codes (described in Section~\ref{sec:codes}) for a range of key atmosphere and surface combinations (Section~\ref{sec: codecompare}). In future work we will expand these comparisons to more surface-atmosphere combinations and to other planets of the Solar System.

\subsubsection{Model Atmospheres and Clouds} \label{sec: Atmos}

All model pixels have vertically heterogeneous atmospheres composed of stacks of horizontally homogeneous and locally plane-parallel layers, which each contain gas molecules and (if desired) cloud particles. The atmosphere is bounded below by a flat, homogeneous surface. Our model atmospheres have a temperature and pressure (T-P) profile representative of a mid-latitude Earth atmosphere as defined by \citet{mcclatchey1972}, divided into 16 total layers (see Figure \ref{fig:tpprofiles}).

We modeled our clouds using MODIS Terra and Aqua satellite data \citep[e.g.,][]{king2004}. We utilized the Level-3 MODIS Atmosphere Daily Global (MADG) Product for 18 May 2013 (i.e., the day of observation for the earthshine measurements from \citet{milespaez2014}) for our model cloud properties. We describe a model cloud by three main properties: its cloud top pressure ($p_{c}$), its optical thickness ($b_{c}$), and its particle size distribution. Our $2\degree \times 2\degree$ model pixels are much larger than typical horizontal variations seen in clouds on Earth, and they are larger than the MODIS pixel size used in the MADG Product. We therefore calculated the average of the three cloud properties of all MODIS pixels within each model pixel and used these average values for the cloud properties of that model pixel.
The MODIS database provides a $p_{c}$ for all of its pixels, and after calculating the average $p_{c}$ for our model pixel, we used our atmospheric T-P profile to determine in which layer the cloud should be placed. All clouds in our model pixels were limited in pressure and temperature so as to avoid mixed-phase clouds or ice clouds.
We calculated the cloud $b_{c}$ for each model pixel by averaging the cloud optical thicknesses for each of the MODIS pixels within our model pixel. The wavelength at which a $b_{c}$ is specified in the MODIS data, $\lambda_{b}$, depends on the surface below the cloud, and we kept this dependency in our model runs. We capped $b_{c}$ at 50 because values above this provided negligible impact on the resulting polarization (see Section \ref{sec: COT}).

\begin{figure}[]
    \centering
    \includegraphics[width=\linewidth]{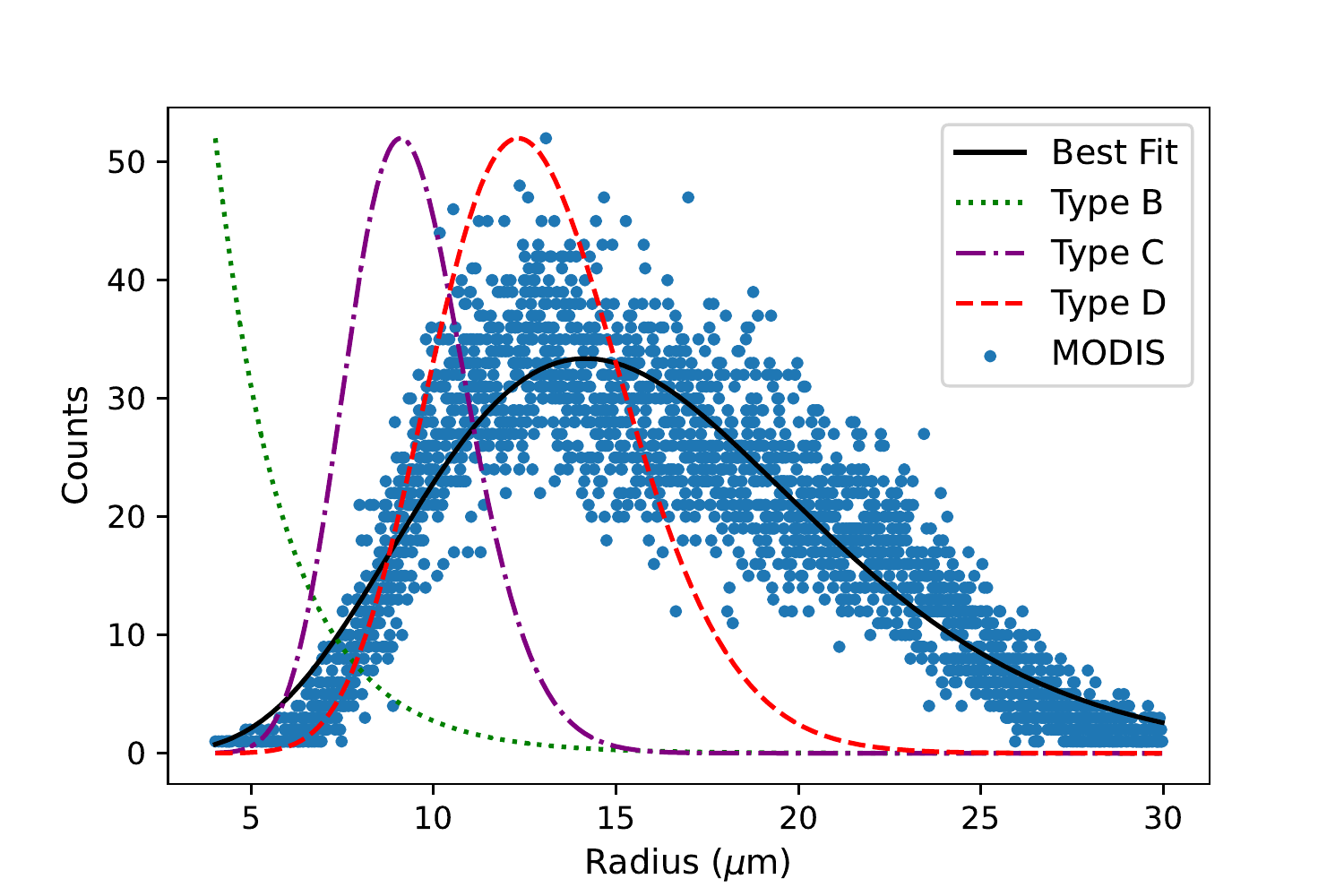}
    \caption{The MODIS liquid water cloud particle sizes (blue dots), as measured by the MADG Product for 18 May 2013 (the day of the earthshine observations). Overplotted are: the best-fit two parameter gamma size distribution (solid black line) and the type B (dotted green line), type C (dashed-dotted purple line) and type D (dashed red line) particle size distributions. The three cloud-type distributions have been normalized in this image to the maximum height of the MODIS data, and they span the majority of the particle sizes measured by MODIS.}
    \label{fig:modisvssize}
\end{figure}

We modeled our cloud particles using Mie theory \citep{derooijvanderstap1984}. Following the MODIS data (Figure~\ref{fig:modisvssize}, blue dots), the model clouds use the two parameter gamma distribution of \citet{hansentravis1974}, which is governed by a particle effective radius $r_{eff}$ in $\mu$m and a dimensionless effective variance $u_{eff}$. Terrestrial liquid water clouds are composed of droplets with radii ranging from $\sim 4 - 5$~$\mu$m up to $\sim 30$~$\mu$m \citep[e.g.,][]{han1994}. The MODIS cloud particle size distribution measured on the day of the earthshine observations is indeed best-matched by a two parameter gamma distribution (Figure~\ref{fig:modisvssize}, solid black line) with radii ranging from $\sim 4$~$\mu$m to $\sim 30$~$\mu$m and a maximum in the distribution at $\sim 14$~$\mu$m.

To analyze the full range of particle radii, we studied three different types of clouds: small droplets with $r_{eff}$=6~$\mu$m, $u_{eff}$=0.4 (type B), medium droplets with $r_{eff}$=10~$\mu$m, $u_{eff}$=0.03 (type C), and large droplets with $r_{eff}$=14~$\mu$m, $u_{eff}$=0.04 (type D). Clouds with larger water droplets require significantly more computing power and are more difficult for \textbf{our codes} to handle. Additionally, the type B clouds are similar to those used by previous theoretical models of the Earth \citep[e.g.,][]{vandiedenhoven2007} as representative of average terrestrial liquid water clouds. Therefore, here we used type B cloud particles in our models. 
In Section~\ref{sec: PSDandAbs} we discuss the effect that the different cloud particle sizes have on our model spectra.

The real part of the complex refractive index of water, $n_w$, is slightly wavelength dependent in the VIS and NIR wavelengths, ranging from 1.344 at $\lambda = 0.4$~$\mu$m to 1.320 at $\lambda = 1.0$~$\mu$m \citep[][]{vandiedenhoven2007, daimonmasumura2007}. The imaginary part of $n_w$ (Im\{$n_w$\}) is small but varies greatly, from $\sim 10^{-8}$ at $0.3$~$\mu$m to $\sim10^{-3}$ at $3$~$\mu$m, with a minimum of $\sim8$ $\times 10^{-10}$ at $0.5$~$\mu$m \citep[][]{segelstein1981, popefry1997}. In Section~\ref{sec: clouds} our models have a wavelength-independent $n_w = 1.335 \pm 0.00001i$ \citep[see][and references therein]{karalidi2011} for simplicity. In Section~\ref{sec: PSDandAbs} we discuss the impact that changing $n_w$ has on our model spectra. Finally, for the comparisons between the two codes in Section~\ref{sec: dapvstar}, we used the wavelength-dependent $n_w$ from \citet[][]{hale1973}.

\subsubsection{Model Surfaces} \label{sec: Surfs}

We used the Level-3 MODIS Yearly Global Land Cover Types (YGLCT) Product for 2013 (i.e., the year of observations for the earthshine measurements) for our model surfaces. This Product identifies 17 total MODIS surface classes, including 11 natural vegetation classes, three human-altered classes, and three non-vegetated classes. As we modeled the Earth-as-an-exoplanet, we condensed these 17 original MODIS surface classes down to five for simplicity: ocean, forest (a combination of deciduous and conifer), grass, sand, and snow and ice. When more than one MODIS surface type existed in our model pixel, we assigned the most abundant MODIS surface type as the surface of our model pixel. 
Our surfaces are modeled as ideal, depolarizing Lambertian surfaces with variable albedos as a function of wavelength. We used the NASA JPL EcoStress Spectral Library \citep[][]{baldridge2009, meerdink2019} to retrieve surface albedo spectra for each surface category\footnote{https://speclib.jpl.nasa.gov}. In a following paper we will study the effect of realistic surfaces by incorporating their Bidirectional Polarization Distribution Functions (BPDF) so that their reflection is non-isotropic.

\section{The Numerical Codes} \label{sec:codes}

\subsection{Flux and Polarization Definitions} \label{sec: defs}

We describe the starlight that is reflected by a planet by the Stokes vector $\pi\textbf{F}$ \citep[see, e.g.,][]{hansentravis1974, hovenier2004}, as

\begin{eqnarray}
\pi\textbf{F}  =  \pi\left[\begin{array}{c} I \\
Q \\
U \\
V\end{array}\right]
\label{eq:fluxvec}
\end{eqnarray}
where parameter $\pi$I is the total flux, parameters $\pi$Q and $\pi$U are the linearly polarized fluxes, and parameter $\pi$V is the circularly polarized flux. All four parameters depend on the wavelength $\lambda$, and their units are in watt per square meter or $W$ $m^{-2} m^{-1}$ when defined per wavelength. Fluxes $\pi$Q and $\pi$U are defined with respect to a reference plane, and in this case we use the planetary scattering plane, which is the plane through the centers of the star, planet, and observer (see top panel of Figure~\ref{fig:plane}).

Based on observations of the Sun \citep[e.g.,][]{kemp1987} and other active and inactive FGK stars \citep[e.g.,][]{cotton2017}, light of a solar-type star can be assumed to be unpolarized when integrated over the stellar disk.
Starlight that has been reflected by a planet is usually polarized because it has been scattered by gases and aerosols or cloud particles in the planetary atmosphere and/or has been reflected by the surface. The total degree of polarization of this light is defined as the ratio of the polarized fluxes to the total flux, or $P_{tot} = \sqrt{Q^{2} + U^{2} + V^{2}} / I$.

Parameter $\pi$V of sunlight that is reflected by an Earth-like planet is expected to be very small \citep[e.g.,][]{hansentravis1974, rossi2018}, and therefore we ignored it in our numerical simulations. Ignoring $\pi$V does not lead to any significant errors in the calculated total and polarized fluxes, as discussed by \citet{stam2005}. $P_{tot}$ can therefore usually be represented by the degree of linear polarization $P = \sqrt{Q^{2} + U^{2}} / I$.

For a planet that is mirror-symmetric with respect to the planetary scattering plane (i.e., the planet is mostly homogeneous), parameter $\pi$U will be effectively zero \citep[e.g.,][]{hovenier1970}. In this case, we define the \textit{signed} degree of linear polarization as

\begin{eqnarray}
P_{s} & = & \frac{-Q}{I},
\label{eq:degofpol}
\end{eqnarray}

The signed degree of linear polarization also indicates the direction of the polarization: if $P_{s} > 0$, the light is polarized perpendicular to the plane containing the incident and scattered light, whereas if $P_{s} < 0$, the light is polarized parallel to the plane.

\subsection{Doubling Adding Program Code} \label{sec: DAP}

The radiative-transfer code DAP uses an efficient adding-doubling algorithm first described by \citet{deHaan1987}, which fully incorporates single and multiple scattering by atmospheric gases as well as aerosol and cloud particles. The adding method allows for the scattering properties of an atmosphere, composed of a stack of individual layers, to be calculated from the scattering properties of the individual  layers, taking into account the repeated reflections at the interfaces between these layers. The doubling method is an extension of the adding method where, if the individual layers are identical, the scattering properties of the combined layer can be obtained through a rapid geometrical doubling manner \citep[see e.g.,][and references therein]{hansentravis1974}.
The code then uses a fast, numerical disk integration algorithm \citep[][]{stam2006} to integrate the reflected light across the planet for all planetary phase angles $\alpha$. DAP has been used to calculate the flux and polarization signals of both terrestrial and gaseous exoplanets \citep[]{stam2006, stam2008, karalidi2011, karalidi2012rainbow, karalidistam2012, treesstam2019, groot2020}.

DAP can model atmospheres of any composition, with as many layers as necessary to describe the full scattering properties of the atmosphere. However, increasing the number of layers results in an increase of the computation cost. Here we used 16 horizontally homogeneous and locally plane-parallel layers to make our calculations tractable, while also capturing the vertical variations of the atmosphere. Our models follow the T-P profile of a mid-latitude Earth \citep[][]{mcclatchey1972}. Each layer contains $H_2O$, $O_3$, $O_2$ and $N_2$ molecules with the volume mixing ratio (VMR) profiles of \citet[][]{mcclatchey1972}. To calculate the absorption properties of these gases, DAP uses the HITRAN 2020 molecular line lists \citep[][]{gordon2022}, and the k-coefficient method. Depending on the MODIS cloud $p_c$ data, atmospheric layers in our models can also contain water cloud droplets of varying $b_c$.

DAP defines the flux vector of stellar light that has been reflected by a spherical planet with radius $r$ at a distance $d$ from the observer (where $d \gg r$) as:

\begin{eqnarray}
\pi\textbf{F}(\lambda, \alpha)  =  \frac{1}{4}\frac{r^2}{d^2}\textbf{S}(\lambda, \alpha)\pi\textbf{F}_0(\lambda)
\label{eq:fluxplanet}
\end{eqnarray}
where $\lambda$ is the wavelength of the light and $\alpha$ is the planetary phase angle, i.e. the angle formed by the star-planet-observer (see top panel of Figure~\ref{fig:plane}). $\pi\textbf{F}_0$ is the flux vector of the incident starlight and $\textbf{S}$ is the $4 \times 4$ planetary scattering matrix with elements $a_{ij}$, which is calculated by DAP \citep[for more information see][]{stam2006}.
For our calculations, we normalized Equation \ref{eq:fluxplanet} assuming $r = 1$ and $d = 1$, and assumed unpolarized incident starlight \citep[e.g.,][]{kemp1987, cotton2017} so that $\pi{F_{0}} = 1$. The total flux reflected by the planet is thus simplified to:

\begin{eqnarray}
\pi{F_{n}}(\lambda, \alpha)  =  \frac{1}{4}a_{11}(\lambda, \alpha)
\label{eq:normalizedflux}
\end{eqnarray}
where $a_{11}$ is the (1,1)-element of the $\textbf{S}$ matrix \citep[see][]{stam2008}, and the subscript $n$ on the flux indicates that it is now normalized. The normalized fluxes $\pi{F_{n}}$ can be straightforwardly scaled for any planetary system using Equation \ref{eq:fluxplanet} and inserting the correct values for $r$, $d$, and $\pi{F_{0}}$. $P_{s}$  (= $-Q / I$) is independent of these values and thus does not require any scaling.

\subsection{Versatile Software for Transfer of Atmospheric Radiation} \label{sec: VSTAR}

VSTAR uses the discrete-ordinate method for the radiative-transfer calculations, and fully incorporates single and multiple scattering by atmospheric gases as well as aerosol and cloud particles. Similar to DAP, VSTAR models the reflected light and polarization phase curves of a planet, in addition to its emission and transmission spectra, for given atmosphere 
and surface properties.
VSTAR utilizes the widely used and validated \citep[see e.g.,][]{kopparla2018} VLIDORT (Vectorized Linear Discrete Ordinate Radiative Transfer) polarized light solution \citep[][]{spurr2006} to solve the polarized, or vector, radiative-transfer equation:

\begin{eqnarray}
\mu\frac{d\textbf{I}_{\nu}(\tau,\mu,\phi)}{d\tau}  =  \textbf{I}_{\nu}(\tau,\mu,\phi) - \textbf{S}_{\nu}(\tau,\mu,\phi),
\label{eq:radtranseq}
\end{eqnarray}
where $\textbf{I}_{\nu}$ is the Stokes vector describing the polarized light at frequency $\nu$  (similar to the Stokes vector of Equation \ref{eq:fluxvec}). $\tau$ is the optical depth of the atmospheric layer and ($\mu$, $\phi$) describe the direction of the light, where $\mu$ is the cosine of the zenith angle and $\phi$ the azimuthal angle. The source function $\textbf{S}_{\nu}$ captures scattering of outside radiation into the beam, thermal emission from the planet, and direct illumination of the atmosphere by an external source. Similar to DAP, VSTAR assumes that this external source is also unpolarized.

VLIDORT replaces the integral that appears in the scattering term of $\textbf{S}_{\nu}$ with a sum using Gaussian quadrature. VSTAR uses a double Gauss scheme in which a separate set of quadrature angles is used for the light beam traveling down through the atmosphere and for the light beam traveling back up through the atmosphere.  
Following the discrete-ordinate method, VSTAR limits the angular distribution of light beams to a discrete number and then sums up these different angular solutions to get the final solution. Increasing the number of quadrature angles (or streams) thus increases the accuracy of the angular representation of the models. However, more streams leads to longer computing time  \citep[for more detailed descriptions and discussions, see][]{bailey2018}. In this study we used 32 streams for our VSTAR models. We confirmed through preliminary tests that no major improvements to the models were acquired with any higher number of streams.

Similar to DAP, VSTAR can model atmospheres of any composition with any number of atmospheric layers. Different atmospheric layers of the VSTAR models, like DAP, can contain water cloud droplets depending on the MODIS cloud data and different VMRs of the $H_2O$, $O_3$, $O_2$ and $N_2$ molecules. However, while DAP uses k-coefficients, VSTAR uses a line-by-line approach to calculate the absorption properties of these atmospheric gases.
As we benchmark VSTAR and DAP against each other (see Sec.~\ref{sec: dapvstar}), all of the VSTAR models copy the same atmospheric structure and surface properties as those used in the DAP models. Finally, both DAP and VSTAR used the 2020 HITRAN molecular line lists for the atmospheric gas absorptions.

\vspace{.2cm}

While DAP and VSTAR follow different approaches to solving the radiative-transfer equation, both codes calculate the resulting reflected light from a planet, and both approaches are historically well validated for vectorized treatments of light \citep[e.g.,][]{spurr2008}. Here we perform the first benchmark of these two approaches for exoplanet modeling against each other and against earthshine data.

\section{Impact of Clouds} \label{sec: clouds}

Clouds on Earth play a significant role in determining the overall polarization state of the planet. Interactions of the incident radiation with cloud particles depolarizes light due to multiple scattering within the clouds, thereby adding more unpolarized background light to the signal and lowering the resulting $P_s$. In this section we use DAP to assess the effects that different properties of water clouds play on the resulting $P_s$ of the reflected light from an Earth-like planet. Unless otherwise stated, the models were generated at a variable resolving power $R$ ranging from $\sim$60 to $\sim$260 in the 0.4 - 1.3~$\mu$m range and $\sim$27 to $\sim$46 in the 1.35 - 2.3~$\mu$m range, with $R \sim 250$ around 1.27~$\mu$m, similar to the resolving power of the binned earthshine data from \citet[][]{milespaez2014}.

\subsection{Effects of Cloud Optical Thickness} \label{sec: COT}

\begin{figure}[ht!]
    \centering
    \includegraphics[width=\linewidth]{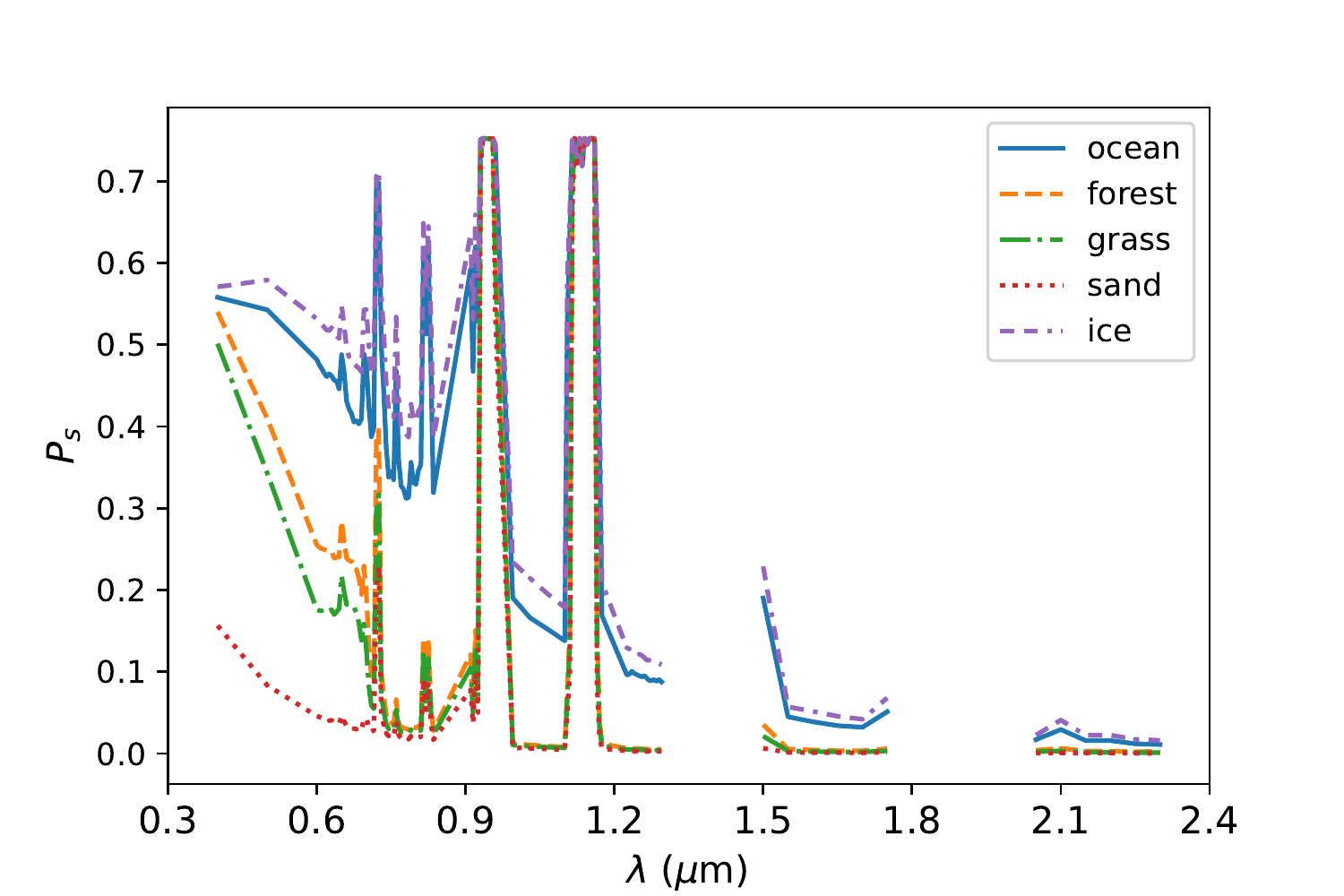}
    \centering
    \includegraphics[width=\linewidth]{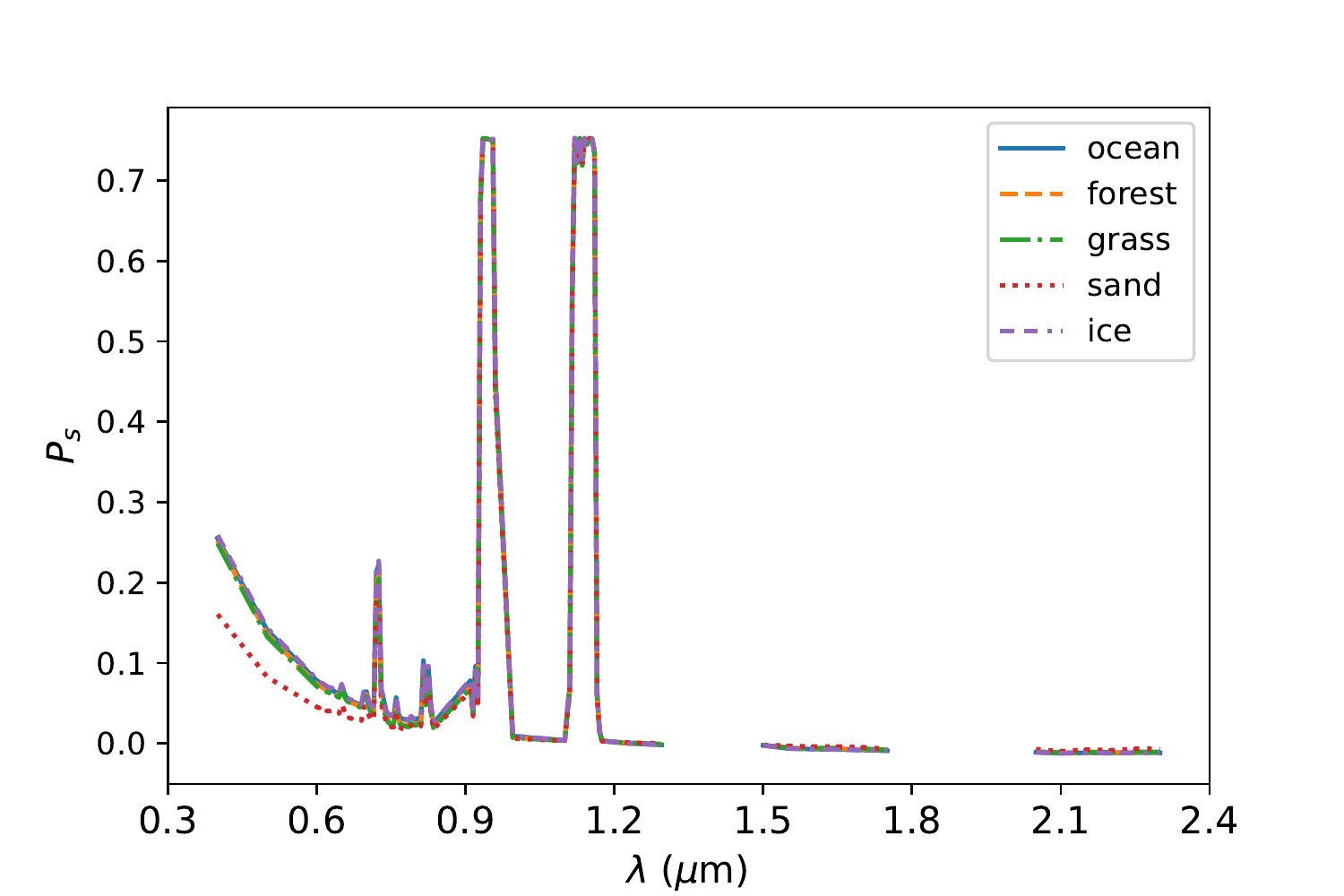}
    \caption{$P_{s}$ as a function of $\lambda$ for $\alpha = 70^{\circ}$ of a planet with an ocean (solid blue line), forest (dashed orange line), grass (dashed-dotted green line), sand (dotted red line), or ice (dashed-dashed-dotted purple line) surface. Our model planets have clear atmospheres (top panel) or atmospheres with one water cloud with $b_c = 8$ and $p_c = 0.710$ bar, consisting of type B particles (bottom panel). The addition of clouds suppresses the surface effects on the resulting spectra.}
    \label{fig:coteffects}
\end{figure}

Figure~\ref{fig:coteffects} shows our model $P_s$ as a function of $\lambda$ for the five different surfaces we consider in our models, and for clear atmospheres (top panel) and cloudy atmospheres (bottom panel). Our cloudy models contain one liquid water cloud layer comprised of type B particles, with $b_c = 8$ and $p_c = 0.710$ bar. All atmospheres contain $O_{2}$ and $H_{2}O$ vapor, which lead to the strong $O_{2}$ A band at 0.76~$\mu$m and the NIR $H_{2}O$ bands near 0.9 and 1.1~$\mu$m. Our cloudless model spectra show clear features of the surfaces, such as the VRE for the forest and grass surfaces around 0.7~$\mu$m. As expected, the introduction of clouds in our model atmosphere reduces, or even erases, the impact of the surfaces on our spectra. The VRE that is easily detectable in the clear atmosphere models (top panel) is reduced in the cloudy models (bottom panel). Additionally, the surface albedos have much smaller effect in the cloudy models, as the spectra for all five surfaces have roughly the same $P_s$ across all wavelengths. These changes are due to the clouds blocking the incoming light and increasing the amount of multiple scattering, thus lowering $P_s$.

\begin{figure}[]
    \centering
    \includegraphics[width=\linewidth]{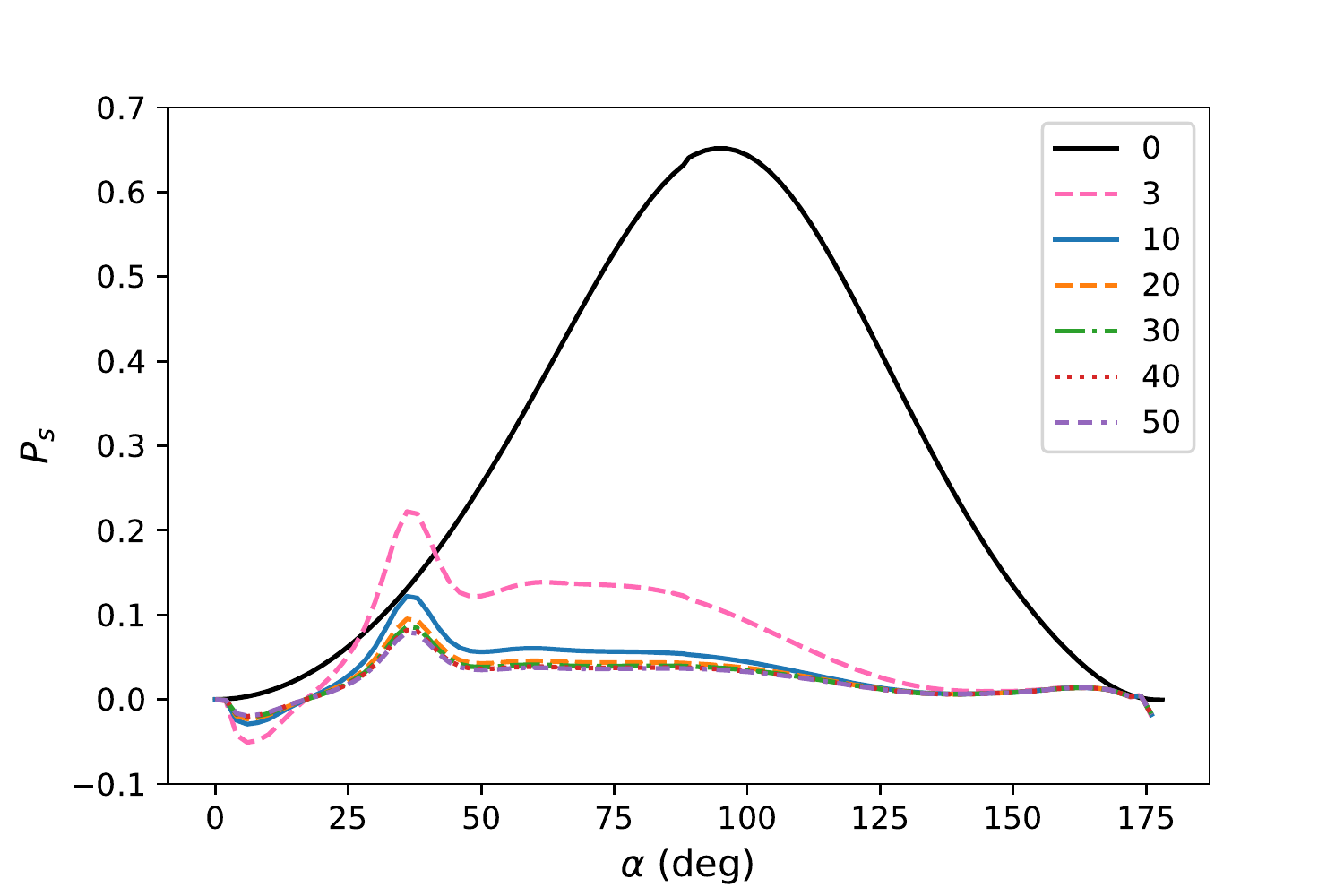}
    \centering
    \includegraphics[width=\linewidth]{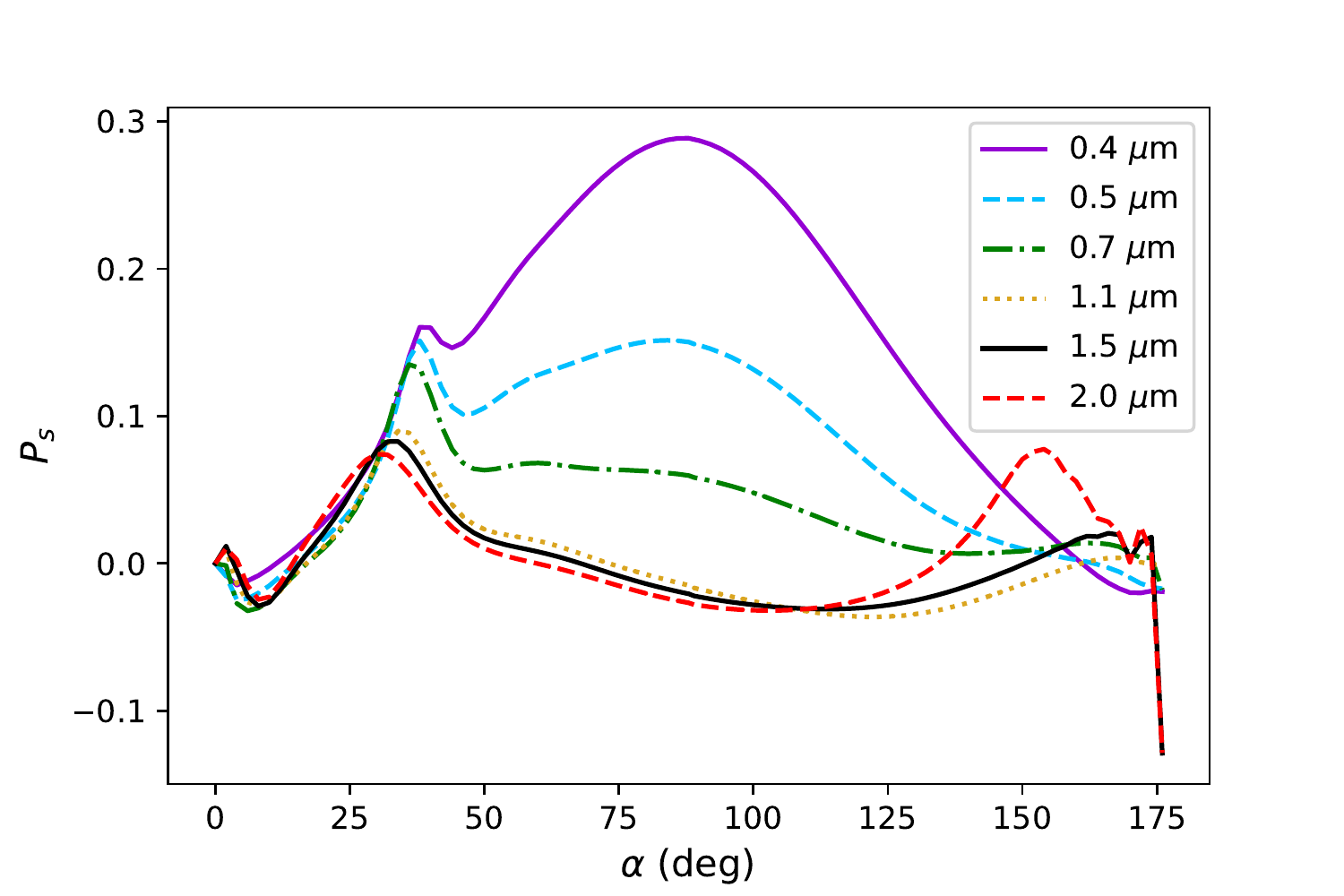}
    \caption{Top panel: the phase dependence of $P_{s}$ for different values of $b_c$ at $\lambda = 0.7$~$\mu$m for planets with an ocean surface and one water cloud layer composed of type B particles with $p_c = 0.710$ bar. The more optically thick the cloud layer, the lower the overall $P_s$. Bottom panel: the wavelength dependence of $P_s(\alpha)$ for liquid water clouds, for a planet with an ocean surface and one cloud layer with $b_c = 8$ and $p_c = 0.710$ bar. As expected, the peak of the rainbow feature ($\alpha \sim 25^{\circ} - 40^{\circ}$) shifts to smaller phase angles at larger wavelengths while the strength of the Rayleigh peak ($\alpha \sim 90^{\circ}$) diminishes.}
    \label{fig:phases}
\end{figure}

For completeness, Figure~\ref{fig:phases} displays $P_s$ as a function of $\alpha$ for cloudy models with: $p_c = 0.710$ bar, $b_c$ ranging from 0 to 50 and at $\lambda = 0.7$~$\mu$m (top panel); and with $p_c = 0.710$ bar, $b_c = 8$ and wavelengths ranging from  0.4 to 2.0~$\mu$m (bottom panel). As expected, the clear atmosphere shows the characteristic bell-shaped curve due to Rayleigh scattering (top panel of Figure~\ref{fig:phases}; solid black line). The addition of clouds decreases $P_s$ across most phase angles and for most $b_c$ values. The only exception is for the $b_c = 3$ model, where the primary rainbow of our liquid water clouds (for $\alpha \sim 30^\circ$ - $40^\circ$) shows a stronger $P_s$ than the Rayleigh scattering. The $P_s$ of our models rapidly converges for $b_c \gtrsim 30$, since $b_c$ larger than this cause the cloud layer to act like a completely depolarizing solid surface and make $P_s$ insensitive to further increases in $b_c$. Any features detected in $P_s$ for $b_c \gtrsim 30$ are due to single, or a few times, scattered light on the upper parts of our model clouds. Finally, the bottom panel of Figure~\ref{fig:phases} shows the wavelength dependence of the rainbow feature for the water clouds, with the location of the rainbow peak shifting to lower $\alpha$ with increasing $\lambda$ \citep[see e.g.,][]{bailey2007, karalidi2012rainbow}.

\subsection{Effects of Particle Size and Cloud Absorption} \label{sec: PSDandAbs}

\begin{figure}[]
    \centering
    \includegraphics[width=\linewidth]{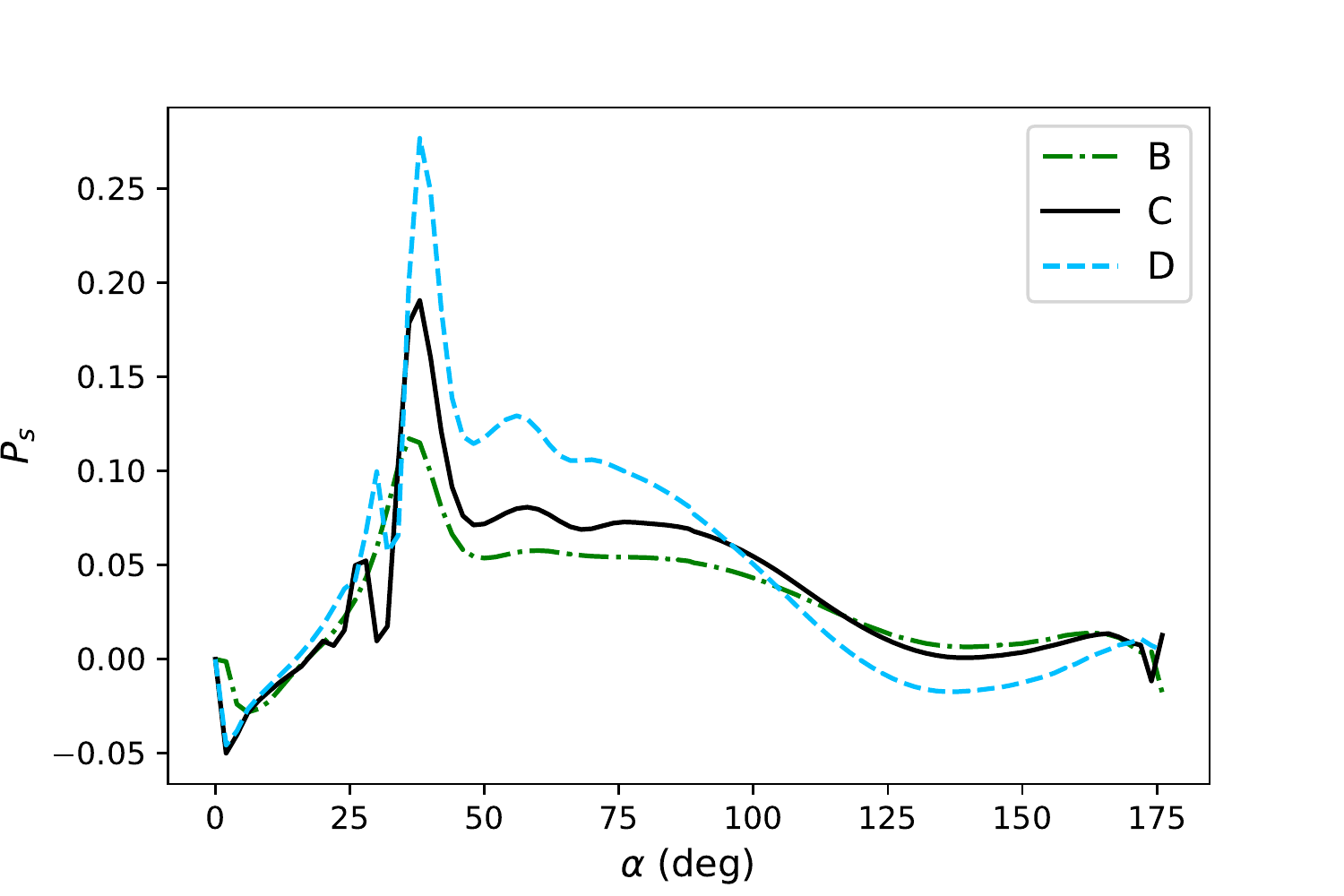}
    \caption{$P_{s}(\alpha)$ at $\lambda$ = 0.7~$\mu$m for the three different types of liquid water clouds tested with DAP, for a planet with a forest surface and a water cloud with $b_c = 10$ and $p_c = 0.710$ bar. The lines represent type B (dashed-dotted green line), type C (solid black line), and type D particles (dashed blue line). Larger particles result in stronger angular features and higher $P_{s}$ across most phase angles.}
    \label{fig:sizephases}
\end{figure}

In Figure~\ref{fig:sizephases} we show $P_s$ as a function of $\alpha$, at $\lambda = 0.7$~$\mu$m, for the three different particle size distributions we explored. These models have a forest surface and a single cloud layer with $b_c = 10$ and $p_c = 0.710$ bar. Our model clouds are simulated using a two parameter gamma distribution with $r_{eff}$ = 6~$\mu$m, $u_{eff}$ = 0.4 (type B); $r_{eff}$ = 10~$\mu$m, $u_{eff}$ = 0.03 (type C); and  $r_{eff}$ = 14~$\mu$m, $u_{eff}$ = 0.04 (type D) (see Section~\ref{sec: Atmos}). Increasing $r_{eff}$ results in higher $|P_s|$, with the larger particles producing stronger angular features than the smaller particles. In particular, the larger particles (types C and D) lead to a stronger and more pronounced primary rainbow than the smaller particles (type B). Additionally, C and D show a pronounced secondary rainbow feature around $\alpha = 60^{\circ}$. 
The differences between the $P_s$ of the three models at the largest phase angles are relatively small because there, most of the reflected starlight has been scattered in the atmospheric layers above the clouds, and therefore the influence from the clouds themselves on the phase curves decreases.

\begin{figure}[]
    \centering
    \includegraphics[width=\linewidth]{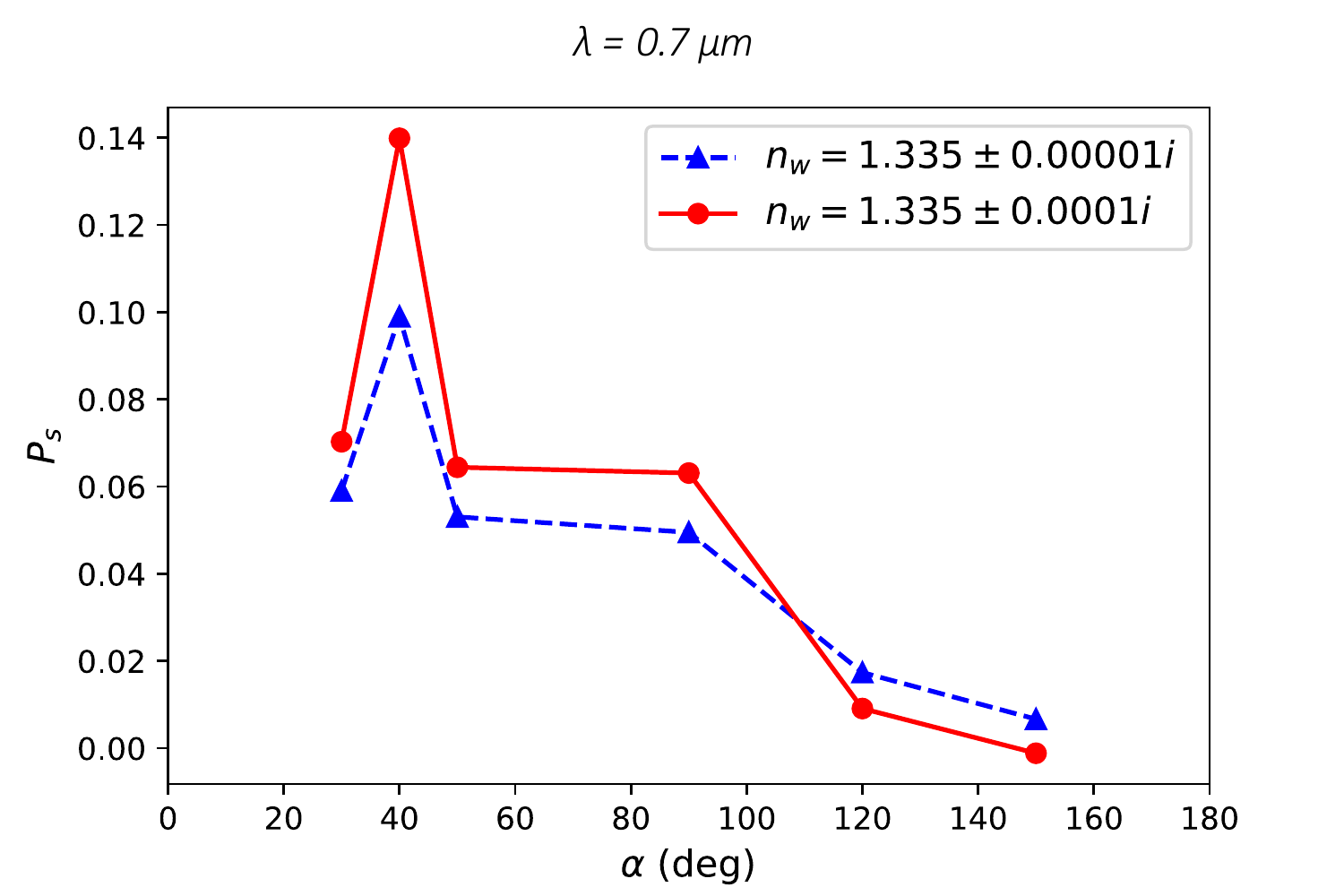}
    \centering
    \includegraphics[width=\linewidth]{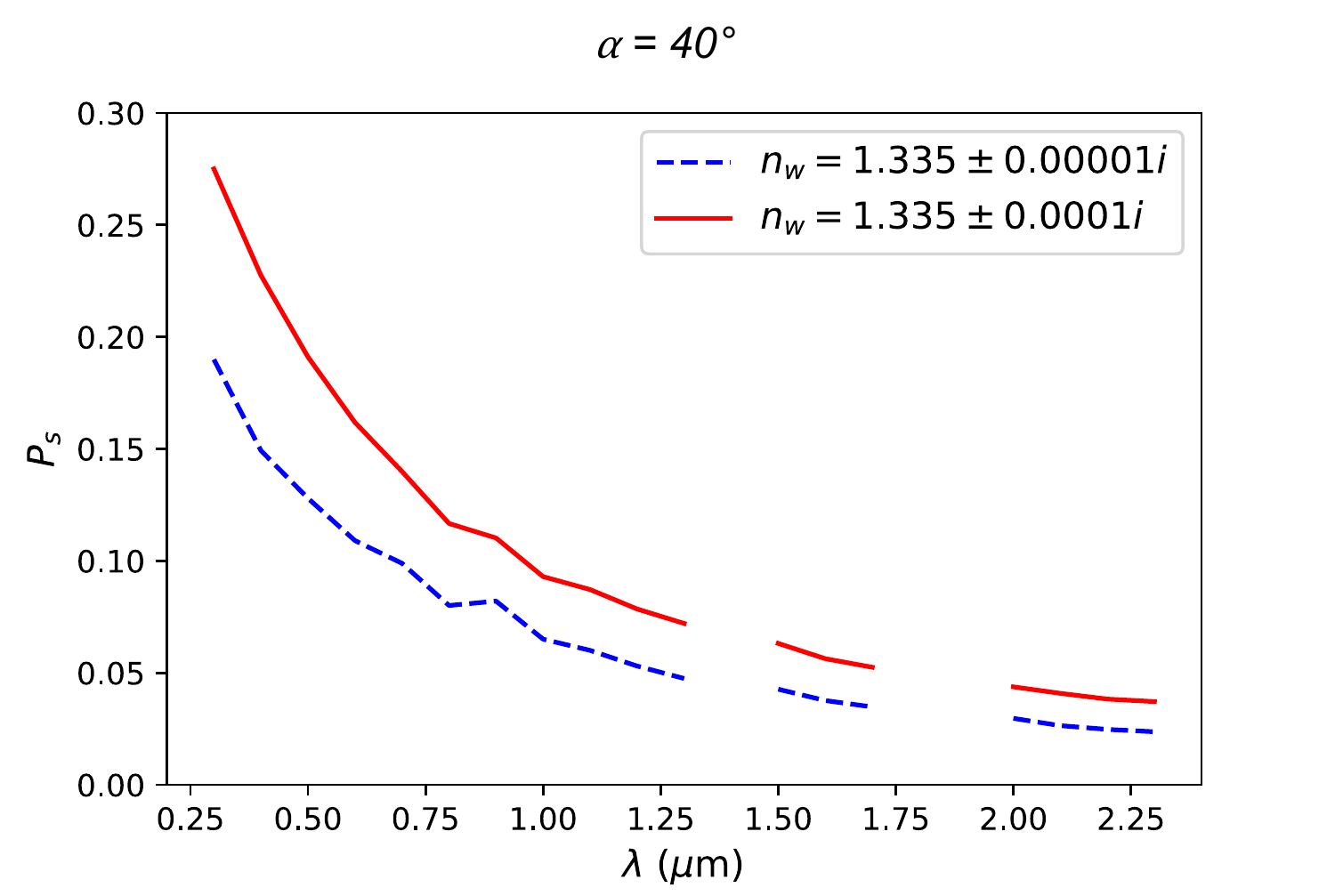}
    \caption{$P_s$ as a function of $\alpha$ (top panel) and $\lambda$ (bottom panel) for our model exoplanet-Earth with water clouds having a constant, wavelength-independent $n_w$. The clouds have either a smaller (Im\{$n_w$\} = 0.00001; dashed blue lines) or a larger absorption (Im\{$n_w$\} = 0.0001; solid red lines). As expected, changing Im\{$n_w$\} affects the total exoplanet $P_s (\alpha)$ and  $P_s (\lambda)$ .}
    \label{fig:indexphase}
\end{figure}

The imaginary part of the complex refractive index of water $n_w$ varies by $\sim$5 orders of magnitude in the VNIR (see Section~\ref{sec: Atmos}). To study the effect of our choice of Im\{$n_w$\} on our model exoplanet-Earth, we also modeled liquid water clouds with a complex refractive index of $1.335 \pm 0.0001i$ so that the imaginary part is one order of magnitude larger than our chosen value. The models in Figure~\ref{fig:indexphase}  incorporate all pixels of our exo-Earth data grid, therefore including clouds with varying $b_c$ and $p_c$ above all five surfaces. These horizontally inhomogeneous spectra were calculated using the weighted sum approximation \citep[see e.g.,][]{stam2008, karalidistam2012}. The flux vector of a planet covered by $M$ different types of pixels is calculated as:  $\pi\textbf{F}(\alpha) = \sum_{m = 1}^{M} w_{m} \pi\textbf{F}_{m}(\alpha)$, where $\pi\textbf{F}_m$ is the flux vector from a single $m$ pixel and $w_{m}$ is the fraction of type $m$ pixels on the inhomogeneous planet, so that $\sum_{m = 1}^{M} w_{m} = 1$.

Figure~\ref{fig:indexphase} (top panel) shows $P_s$ of our model exoplanet-Earth as a function of $\alpha$ at $\lambda = 0.7$~$\mu$m, for the two different values of $n_w$. To reduce computation time we only ran these models at six key phase angles. The angles were chosen to highlight the primary rainbow feature around $\alpha=40^{\circ}$, the planet at its furthest distance from its star at $\alpha = 90^{\circ}$, and a few larger phase angles in order to approximate the full shape of the phase dependence.
For most of the phase angles, 
the model with the larger imaginary part of $n_w$ (red line) produced higher levels of polarization than the model with the smaller imaginary part of $n_w$ (blue line), due to more absorption of the light through the water particles. This higher absorption leads to less multiple scattering of the light by the clouds and therefore less depolarization of the light. As we approach limb-viewing geometry at larger $\alpha$, most of the reflected starlight gets scattered by the atmospheric gases in the layers above the clouds, and light that does end up reaching the cloud layers is more likely to be absorbed rather than scattered by the water droplets with the greater imaginary part of $n_w$. This leads to an overall lower level of polarization than the particles with smaller imaginary part of $n_w$.

Finally, the bottom panel of Figure~\ref{fig:indexphase} shows $P_s (\lambda)$ for the two different $n_w$ at $\alpha = 40^{\circ}$, where the polarization of the liquid water clouds is strongest due to the primary rainbow feature. Here we are not interested in the spectral features of the atmosphere, rather just the change in the polarization continua due to the different Im\{$n_w$\}. Thus, these spectra are broad band, with a constant spectral resolution of 0.1 $\mu$m. As expected, the model with the larger Im\{$n_w$\} (red line) shows higher $P_s$ across the entire spectrum. 

\begin{figure*}[ht!]
    \centering
    \includegraphics[width=18cm]{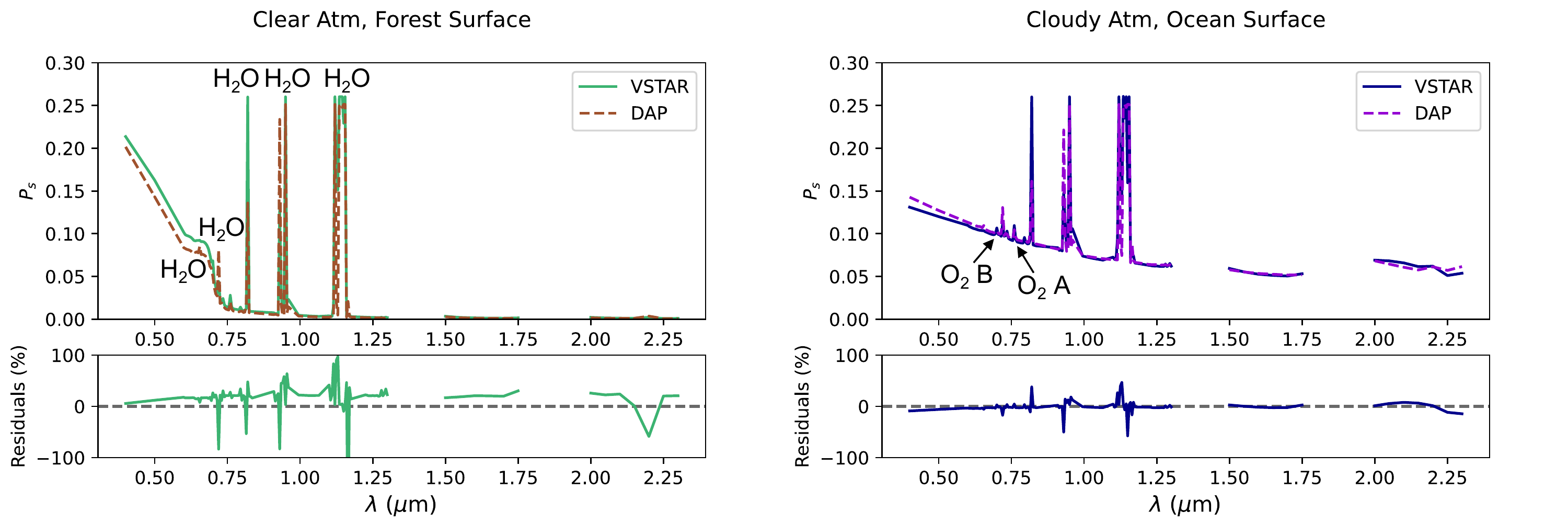}
    \caption{$P_s (\lambda)$ for VSTAR (solid lines) and DAP (dashed lines) models at $\alpha = 40^{\circ}$. Our models have a forest surface and clear atmosphere (left panel) or an ocean surface and one water cloud layer with $b_c = 10$ and $p_c = 0.710$ bar, consisting of type B particles with a wavelength-dependent $n_w$ (right panel). The major $H_2O$ spectral features are labeled in the clear forest $P_s$ plot (left panel), while the major $O_2$ spectral features are labeled in the cloudy ocean $P_s$ plot (right plot). The differences seen in the absorption lines is due to the different approaches DAP and VSTAR use to model absorption.}
    \label{fig:dapvsvstar}
\end{figure*}

\section{Model Comparisons} \label{sec: codecompare}

\subsection{Correcting for Lunar Depolarization} \label{sec: depol}

To use earthshine observations as a benchmark for theoretical models we need to correct our observations for the lunar depolarization. Reflection off the Moon results in a depolarization of the polarized earthshine. This is due to the rough particulate regolith of the Moon, which induces high levels of back-scattering. 
\citet{dollfus1957} showed that the bright regions on the Moon depolarize the continuum of the incident light more strongly than the dark regions, due to the different porosity levels and particulate radii of the lunar soil in mare as compared to highlands and craters. \citet{dollfus1957} also noted a wavelength dependence of the depolarization for visible wavelengths, at a factor of approximately $3.3\lambda/550$ ($\lambda$ in nm).

In general terms, the lunar depolarization factor is defined as: 
$\epsilon  =  P^{out} / P^{in}$,
where $P^{out}$ is the $P$ of the light reflected by the Moon (i.e., the earthshine) and $P^{in}$ is the $P$ that is incident on the Moon (i.e., the true polarization of the Earth). 
To date no direct measurements of $\epsilon$ exist. Therefore, here we follow the discussion of \citet{bazzon2013} and assume that

\begin{equation}
\log \epsilon(\lambda, a_{603}) = -0.61 \log a_{603} - 0.291 \log \lambda - 0.955 
\label{eq:logeps}
\end{equation}
where $\log a_{603}$ is the lunar albedo at 603 nm, and the wavelength $\log \lambda$ is measured in $\mu$m. This method was validated in previous studies of earthshine polarimetry by \citet[][]{sterzik2019, sterzik2020}. We acknowledge that this equation from \citet[][]{bazzon2013} was determined based on their observations at visible wavelengths only, but at this time we are not aware of any study providing a depolarization factor for the NIR.

Due to the small length of the slit at the NOT used for the earthshine observations, \citet{milespaez2014} had to alternate exposures of the Moon and sky positions throughout their observing night. The optical part of their measured spectrum is thus an average from multiple regions of the lunar surface. Therefore, for our calculations of $\epsilon$, we use an average lunar albedo of $a_{603} = 0.1359$ \citep[based on measurements by][]{velikodsky2011}. After calculating $\epsilon$ through Equation~\ref{eq:logeps}, the corrected, true polarization of the Earth is then calculated through: $P^{in} = P^{out} / \epsilon$.

\subsection{DAP versus VSTAR} \label{sec: dapvstar}

This paper is part of a larger project aimed at comparing DAP and VSTAR. Here we compared the two codes to confirm that the results for our exoplanet-Earth are independent of the code we used and thus the two different solutions to the radiative-transfer equation. In future work we will provide more in-depth comparisons, incorporating more surface-atmosphere interactions and comparing to other planets in the Solar System.

In Figure~\ref{fig:dapvsvstar} we compare $P_s (\lambda)$ for DAP (dashed lines) and VSTAR (solid lines), at a phase angle of $\alpha = 40^{\circ}$.  
The models were generated at constant spectral resolutions of 5 nm for $\lambda = 0.4 - 1.3$~$\mu$m (R $\sim$ 250 at $\lambda = 1.27$~$\mu$m) and 50 nm for $\lambda = 1.3 - 2.3$~$\mu$m (R $\sim$ 32 at $\lambda = 1.6$~$\mu$m), since we are only interested in the level of the continua and not the spectral features at the longest wavelengths. The left panel shows models for clear atmospheres above a forest surface. The right panel shows models with an ocean surface and atmospheres that contain one liquid water cloud layer with $b_c = 10$ and $p_c = 0.710$ bar, consisting of type B particles with the wavelength-dependent $n_w$ of \citet[][]{hale1973}. Also shown are the residuals between the two codes, calculated as the relative percent change in the VSTAR spectra with respect to the DAP spectra.

\begin{figure*}[ht!]
    \centering
    \includegraphics[width=18cm]{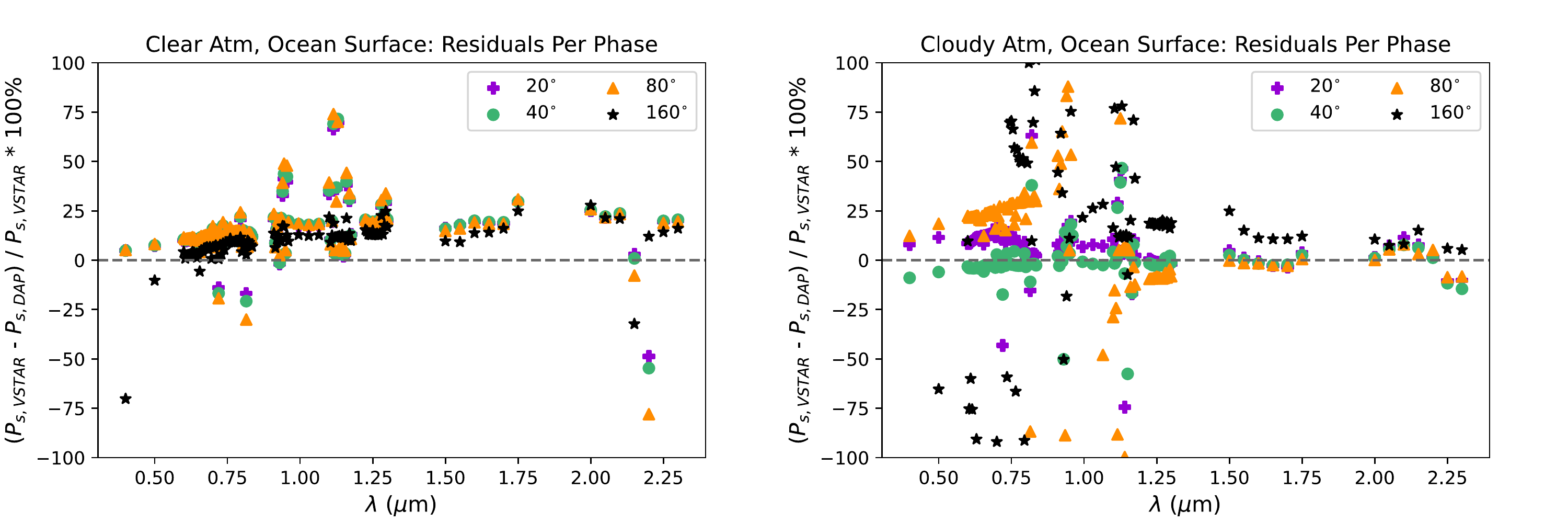}
    \caption{Residuals between the VSTAR and DAP models as a function of $\lambda$ for $\alpha$'s of $20^{\circ}$ (purple pluses), $40^{\circ}$ (green circles), $80^{\circ}$ (orange triangles), and $160^{\circ}$ (black stars). All models have an ocean surface and  either clear atmospheres (left panel) or atmospheres containing one liquid water cloud layer similar to Figure~\ref{fig:dapvsvstar} (right panel). The residuals show an $\alpha$ dependence, with the lower residuals noted at the primary rainbow feature ($\sim40^{\circ}$).}
    \label{fig:diskint}
\end{figure*}

The continua of the models for both cases match well across the entire spectral range, with the majority of differences appearing around the $O_2$ and $H_2O$ absorption bands (see residuals). Both codes use the HITRAN 2020 molecular line lists \citep[][]{gordon2022} to calculate the $O_2$ and $H_2O$ absorption, but VSTAR uses line-by-line and DAP k-coefficients for these calculations. The small discrepancies in the peaks and widths of the absorption features between the models are due to the wavelength grid that the k-coefficients were made with, which lost some of the finer detail of the absorption lines compared to VSTAR's line-by-line approach.
The larger residuals of the NIR continuum for the clear forest comparisons is due to the low polarization of these models at longer wavelengths, where slight differences in $P_s$ leads to large relative percent change. However, the two models still match well across this continuum ($P_s$ = 0.38\% (0.32\%) for VSTAR (DAP) at 1.03 $\mu$m; $P_s$ = 0.15\% (0.13\%) for VSTAR (DAP) at 1.6 $\mu$m).

Figure~\ref{fig:diskint} shows the residuals between the VSTAR and DAP models for four different $\alpha$'s ranging from $20^{\circ}$ to $160^{\circ}$.
All models now have an ocean surface and atmospheres that are either clear (left panel) or with one liquid water cloud layer similar to Figure~\ref{fig:dapvsvstar} (right panel). The VSTAR-DAP residuals exhibit a phase angle dependence. For the cloudy models (right panel), the minimum residuals at $\alpha = 40^{\circ}$ (i.e., near the primary rainbow feature) are due to the strong rainbow polarization of the liquid water clouds dominating $P_s$.
The maximum residuals at $\alpha = 160^{\circ}$ were also observed in \citet{karalidistam2012} and, similar to that study, are most likely caused by the different disk-integration methods used in VSTAR and DAP.
Finally, in both the clear and cloudy models, the greatest residuals occur in the NIR $H_2O$ absorption bands (49\% (clear) and 83\% (cloudy) at 0.93 $\mu$m; 74\% (clear) and -24\% (cloudy) at 1.12 $\mu$m; versus 19\% (clear) and -1.6\% (cloudy) at 1.6 $\mu$m (the continuum) for the $\alpha = 80^{\circ}$ comparisons)
and are due to the wavelength grid with which the k-coefficients were created.

All VSTAR and DAP models compared here used the same (Lambertian) surface treatment and had the same atmospheric structure, including number of atmospheric layers as well as T-P, composition, and (when applicable) cloud profiles. Both codes followed the derivations of \citet[][]{peck1972} for Rayleigh scattering in dry air and modeled clouds using Mie theory. Finally, both codes utilized the HITRAN 2020 database for the molecular absorption, with VSTAR using line-by-line and DAP k-coefficient calculations. While some differences between the VSTAR and DAP models can be attributed to the different absorption calculation methods and disk-integration methods, a number of discrepancies exist that we cannot currently identify the source of. We will address these discrepancies in future work.

\begin{figure*}[ht!]
    \centering
    \includegraphics[width=18cm]{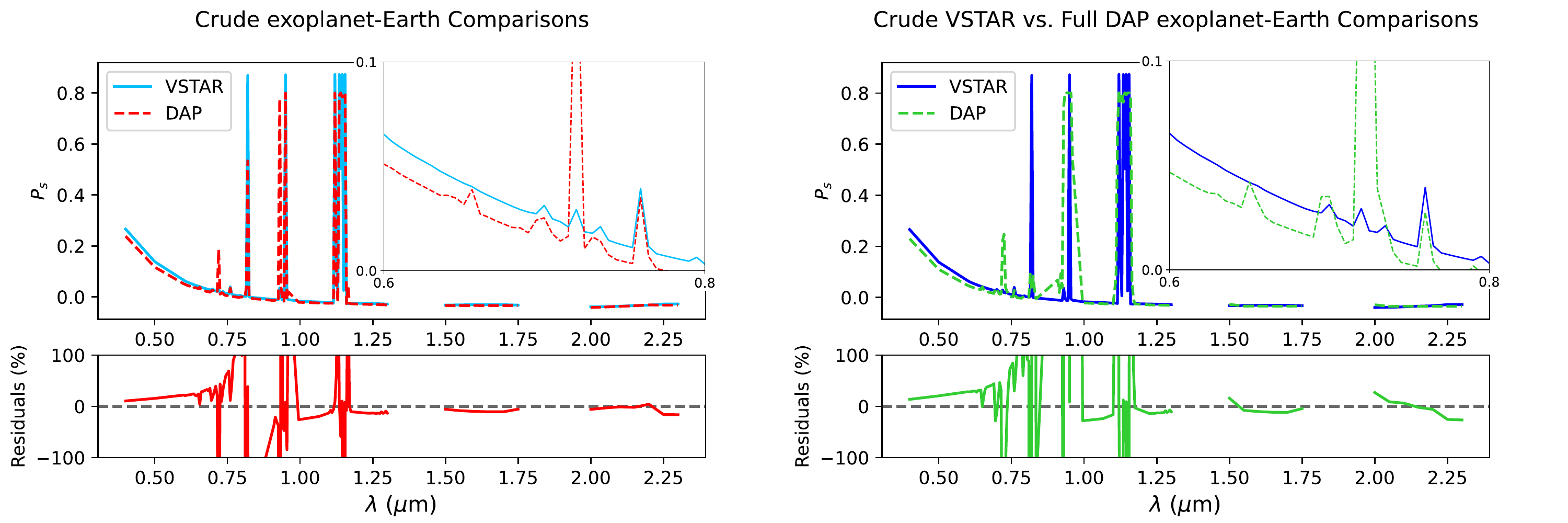}
    \caption{Comparison between blended models of DAP (dashed lines) and VSTAR (solid lines) for $P_s$ as a function of $\lambda$. The left panel shows crude exoplanet-Earth models for both codes generated from blending four simple cloudy and clear ocean and forest models. The right panel displays the crude VSTAR exoplanet-Earth model compared against the full-blown DAP exoplanet-Earth model, which takes into account all five surface types and all clouds with type B particles (with a wavelength-independent $n_w$) and varying $b_c$ and $p_c$. These models were created at $\alpha = 106^{\circ}$ to match the geometry of the Earth during the earthshine observations. Residuals are shown for both comparisons. The inset plot of each panel shows a zoomed-in view of the spectra to highlight the main differences around the VRE (centered at $\lambda \sim 0.7$ $\mu$m).} 
    \label{fig:blendcomp}
\end{figure*}

\subsection{Earthshine versus Models} \label{sec: earthcompare}

In this section we compare different models of our exoplanet-Earth from both DAP and VSTAR, at a phase angle of $\alpha = 106^{\circ}$, to each other and to both the original earthshine observations as well as the observations corrected for the lunar depolarization.

In Figure~\ref{fig:blendcomp} we compare $P_s$ of exoplanet-Earth models generated by DAP and VSTAR. Due to computational time restrictions, the exoplanet-Earth VSTAR models (solid lines) represent a crude blending of four models: the two models shown in Figure~\ref{fig:dapvsvstar}; a model with an ocean surface and a clear atmosphere; and a model with a forest surface and the same cloud as in Figure~\ref{fig:dapvsvstar}. For pixels in the MODIS data with ocean or ice surfaces we used the VSTAR ocean models, while the VSTAR forest models were used for all other surface types. The cloudy VSTAR models were used for pixels containing clouds with $b_c \geq 3$ (and any $p_c$), and the clear VSTAR models were used for all other pixels.  
The DAP model was made by blending just the four models used in the VSTAR model (left panel; dashed red line), or all the surface and cloud models included in the MODIS data [but now for water cloud particles with a wavelength-independent $n_w$ of $1.335 \pm 0.00001i$, due to computational time restrictions] (right panel; dashed green line). This full exoplanet-Earth DAP model in the right panel was generated using the same weighted sum approximation as described in Section~\ref{sec: PSDandAbs}.
All models had spectral resolutions of 5 nm for $\lambda = 0.4 - 1.3$~$\mu$m and 50 nm for $\lambda = 1.3 - 2.3$~$\mu$m, resulting in $R \sim 250$ at $\lambda = 1.27$~$\mu$m. The residuals of the two comparisons are shown in the bottom of each panel, calculated as the relative percent change in the VSTAR spectra with respect to the DAP spectra.

The general shape and slope of $P_s$($\lambda$) match very well between the DAP and VSTAR models. The continua of the spectra overlap across most wavelengths, and the differences in the large $H_2O$ spectral features are expected due to the different methods of handling absorptions between the two codes (line-by-line for VSTAR vs. k-coefficients for DAP). The inclusion of extra surfaces in the full-blown DAP exoplanet-Earth model (right panel, dashed green line) leads to more differences between the DAP and the VSTAR models. In particular, the largest variations occur around the VRE ($\lambda \sim 0.7$ $\mu$m), with $\delta P_s$ $(= P_{s,VSTAR} - P_{s, DAP}) \sim 0.009$ for the crude exoplanet-Earth model comparison (left panel, inset plot) and $\delta P_s \sim 0.012$ for the crude VSTAR model compared to the full DAP model (right panel, inset plot). This is because the crude VSTAR exoplanet-Earth model only includes forest surfaces for its land pixels, while the full-blown DAP exoplanet-Earth model replaces some of these vegetated surfaces with sand or ice surfaces depending on the specific pixel in the data grid, thereby slightly decreasing the strength of the VRE. The small differences at the longer wavelengths in the right panel can be attributed to the different $n_w$ used for the liquid water clouds, which can lead to slight differences in the absorptive properties of the cloud particles and therefore alter the resulting $P_s$ (see Section~\ref{sec: PSDandAbs}).

\begin{figure}[]
    \centering
    \includegraphics[width=\linewidth]{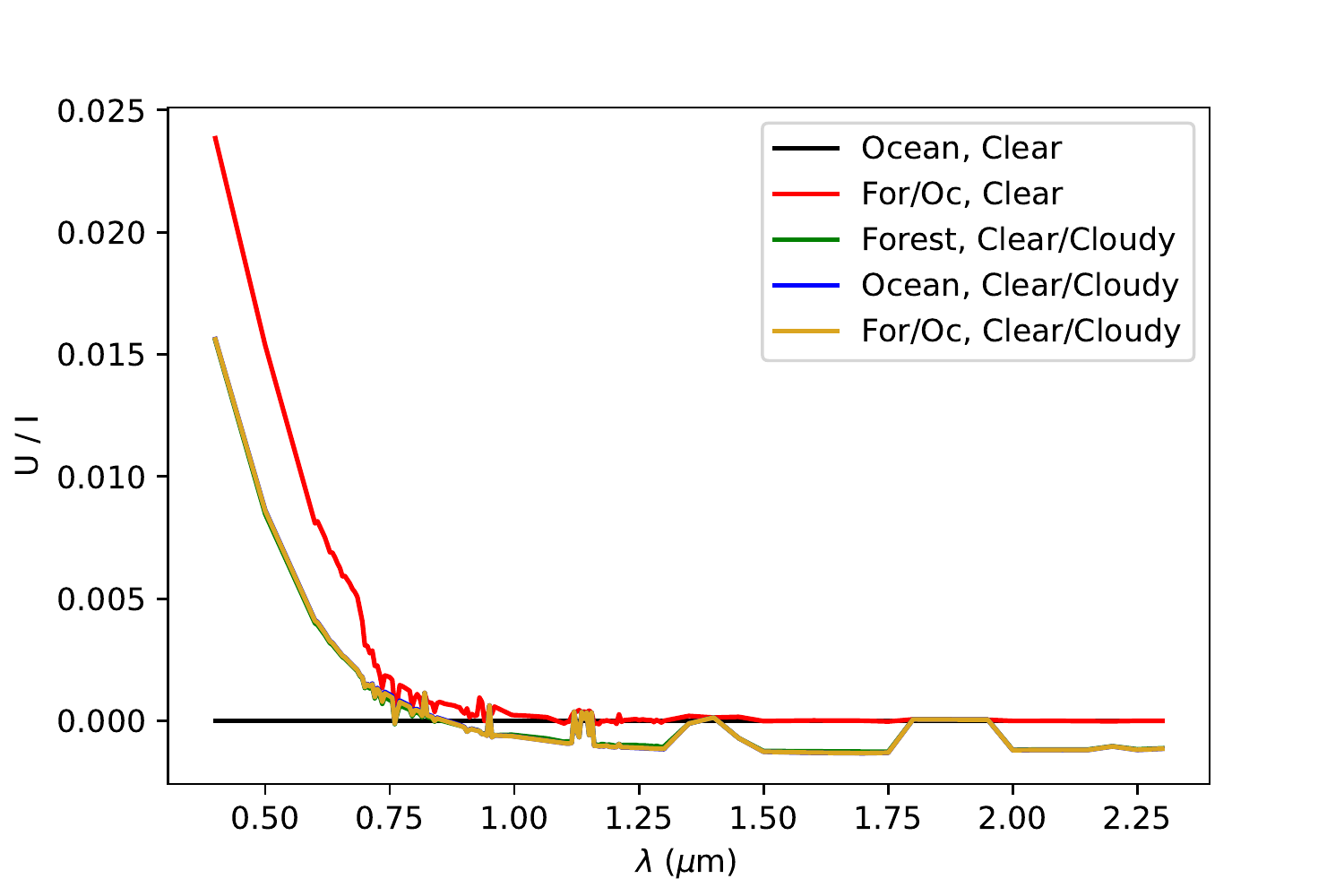}
    \centering
    \includegraphics[width=\linewidth]{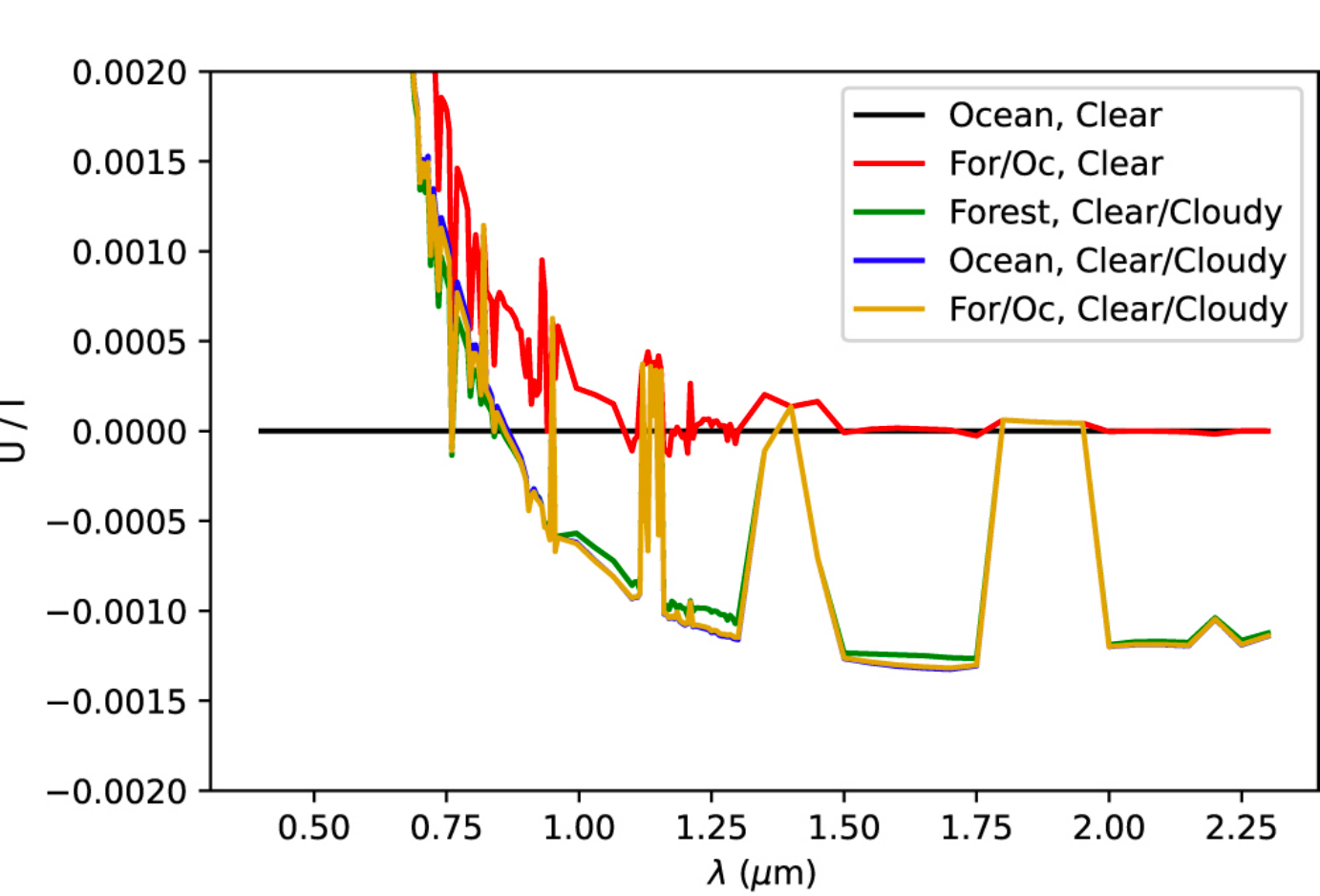}
    \caption{Comparisons between the ratios of Stokes U to Stokes I for a fully homogeneous planet (black line) and planets with different heterogeneities generated by VSTAR. All models are for $\alpha = 106^{\circ}$. The homogeneous model is for a clear atmosphere above an ocean surface, whereas the heterogeneous models show the resulting U/I for crude exoplanet-Earth models created from different combinations of the four simple VSTAR models from Figure \ref{fig:blendcomp}. The bottom panel shows the spectra with a zoomed-in view on the y-axis to better visualize their structures at longer wavelengths.}
    \label{fig:homovshetero}
\end{figure}

It is important to note that these spectra show $P_s$~$(= -Q/I)$, so they hold information about the direction of $P$ of the reflected light (Section~\ref{sec: defs}). For the given $\alpha_\earth$ at the time of the observations, $P_{s,\mathrm{DAP}}$ is low, with the continuum becoming negative at $\lambda \gtrsim 0.75$~$\mu$m, indicating that the direction of the polarization has flipped to being parallel rather than perpendicular to the scattering plane. This result is in line with previous studies and is due to a combination of the physical and micro-physical properties of the clouds used in our models \citep[e.g.,][]{stam2008}. Analyzing the angles of the polarization of the crude heterogeneous VSTAR model, we find a small flip in the angles at the same wavelength as the DAP model, with the VSTAR angles tending towards 0$\degree$. 
Finally, we note that for the VSTAR spectrum in Figure~\ref{fig:blendcomp} we plot the ratio of Stokes Q to Stokes I ($-Q/I$) in order to provide a more direct comparison between the VSTAR and DAP models. This is due to the weighted sum approximation used for our full exoplanet-Earth DAP model, which results in the Stokes U being zero. On the other hand, the output from VSTAR can be used to produce a nonzero Stokes U. In future papers we will use the method of \citet{karalidi2012rainbow} to produce realistic heterogeneous models ($U \neq 0$) for DAP and compare against the full heterogeneous VSTAR models (taking into account all surfaces and cloud variations). 

The effect of planet heterogeneity on Stokes U can be seen in Figure~\ref{fig:homovshetero}, where we plot the ratio of linearly polarized flux $\pi$U to total flux $\pi$I for different models generated by VSTAR. These include a homogeneous ocean planet with a cloud-free atmosphere (black line) along with four crude exoplanet-Earth models generated using different combinations of the simple VSTAR models from Figure~\ref{fig:blendcomp}: a mixed surface planet with forest and ocean surfaces and a cloud-free atmosphere (red line); a combination of clear and cloudy forest models (green line); a combination of clear and cloudy ocean models (blue line); and the crude VSTAR exoplanet-Earth model of Figure~\ref{fig:blendcomp} that incorporates both surfaces and clear and cloudy models (gold line). All four heterogeneous cases are mapped to the full exoplanet-Earth data grid (Section~\ref{sec: MODIS}) using a similar method as the crude VSTAR exoplanet-Earth model of Figure~\ref{fig:blendcomp}. As expected, the homogeneous case has $U/I \sim 0$ at all wavelengths due to the disk symmetry. The introduction of heterogeneity breaks this symmetry and results in $U/I \neq 0$, with the cloud heterogeneity (green and blue lines) affecting $U/I$ more than the surface heterogeneity (red line). Note that the spectrum of the homogeneous ocean planet with the clear and cloudy atmospheres (blue line) is almost completely covered by the spectrum of the full-blown crude VSTAR exoplanet-Earth model (gold line). This is due to the strong contribution of the Atlantic Ocean on our exoplanet-Earth data grid (see Figure~\ref{fig:Moonloc}), which has a stronger effect on the resulting polarization than the forest surface pixels.  

\begin{figure*}[]
    \centering
    \includegraphics[width=18cm]{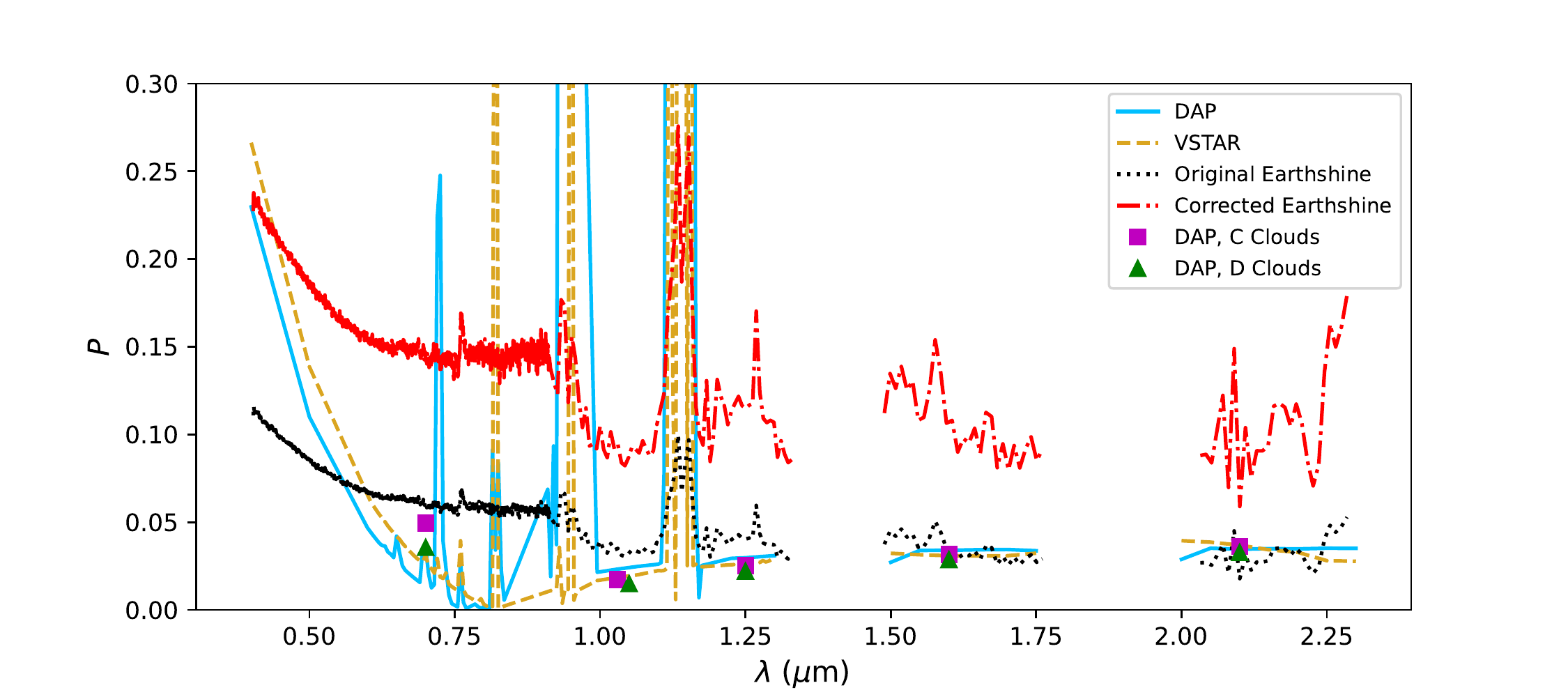}
    \caption{Comparison between the full-blown exoplanet-Earth model generated by DAP (solid blue line) and a crude exoplanet-Earth model created from the four simple VSTAR models from Figure \ref{fig:blendcomp} (dashed gold line) with the original earthshine observations (dotted black line) and the earthshine observations corrected for lunar depolarization (dashed-dotted red line). The DAP and VSTAR models are at $\alpha = 106^{\circ}$ to match the observations. The DAP model includes all five surface types and clouds with varying $b_c$ and $p_c$, while the VSTAR model includes only forest and ocean surfaces and clouds with a set $b_c$ and $p_c$. The DAP and VSTAR model clouds are composed of type B particles. The VSTAR cloud particles used the wavelength-dependent $n_w$ of \citet[][]{hale1973}, while the DAP cloud particles used a wavelength-independent $n_w$ due to computational time restrictions. Also included in this plot are data points from two additional crude exoplanet-Earth models created by DAP, but now using the type C (purple squares) and type D (green triangles) cloud particles with the same wavelength-independent $n_w$ as the type B clouds of the DAP model.}
    \label{fig:earthcompare}
\end{figure*}

Figure~\ref{fig:earthcompare} displays comparisons between the exoplanet-Earth models generated by DAP (solid blue line) and VSTAR (dashed gold line) against both the original earthshine observations (dotted black line) and the observations corrected for lunar depolarization (dashed-dotted red line). The VSTAR model was generated from the same four simple models used to generate the crude VSTAR exoplanet-Earth model (Figure~\ref{fig:blendcomp}), but now incorporates Stokes U as well. Since $U = 0$ for the DAP data, the DAP model is simply taken to be the absolute value ($P_{DAP} = \sqrt{Q^2} / I$) of the DAP spectrum from the right panel of Figure~\ref{fig:blendcomp} for the full-blown DAP exoplanet-Earth. Again, the models had spectral resolutions of 5 nm for $\lambda = 0.4 - 1.3$~$\mu$m and 50 nm for $\lambda = 1.3 - 2.3$~$\mu$m, resulting in $R \sim 250$ at $\lambda = 1.27$~$\mu$m. The earthshine observations were originally measured at $R \sim 690$ in the NIR but the spectrum was binned down to increase the S/N, resulting in $R \sim 208$ at $\lambda = 1.27$~$\mu$m.

Comparing the DAP and VSTAR models in Figure~\ref{fig:earthcompare}, we see similar trends as those discussed in previous sections. The shape and slope of the models match each other well, with only slight differences due to the different treatments of the atmospheric absorptions between the two codes, as well as the inclusion of more surfaces in the DAP model. Additionally, the inclusion of Stokes U in the VSTAR spectrum may add to the slight increase in $P$ in the shorter wavelengths compared to the DAP spectrum (see Figure~\ref{fig:homovshetero}). As expected, correcting the earthshine for lunar depolarization (Section~\ref{sec: depol}) resulted in an increase of $P$ by $\sim$ 4 - 12$\%$ (depending on $\lambda$), with $\delta P$ reaching $\sim 18\%$ in the 1.12~$\mu$m NIR $H_2O$ band.

Interestingly, the spectral slope of the earthshine observations differs from that of the models. While both models agree with one another, they both underestimate the polarization of the earthshine, especially after correcting for lunar depolarization. In an attempt to find a realistic value for the lunar albedo that would improve the fit between our models and the corrected earthshine, we tested different values for $a_{603}$ (see Equation~\ref{eq:logeps}). However, while lower values of $a_{603}$ decreased the resulting $P$ of the corrected earthshine, no value provided a match between the models and the observed spectra. We therefore opted to stick with the average value of $a_{603}$ = 0.1359. Both of the models and the observations show a Rayleigh-like slope at the shorter wavelengths, and the $P$ of the models match the $P$ of the corrected earthshine at $\sim$ 0.4 $\mu$m. However, the earthshine observations are much flatter than the model spectra across all wavelengths.

We note that in both the original and corrected earthshine observations, there is an offset in the level of $P$ between the reddest wavelengths of the optical spectrum ($\lambda \gtrsim 0.8$ $\mu$m) and the NIR continuum. During their observations, \citet[][]{milespaez2014} collected the optical and NIR data simultaneously, but were unable to point the two telescopes at exactly the same region of the Moon. The offset is therefore not expected to be real, but instead a product of observing different lunar areas with different properties.

Changes in the spectral slope at shorter wavelengths are connected with the existence and properties of clouds in an atmosphere. Additionally, as discussed in Section~\ref{sec: PSDandAbs}, changes in the micro- and macro-physical properties of the clouds affect the resulting $P$ in the NIR. The difference between our models and the observations indicate that our approach of using just a single cloud layer for a given pixel, with one type of model cloud (type B) across all pixels, is over-simplified and we would need a broader range of cloud properties to match the observations. Addressing this issue will be part of future work. Additionally, the flatter spectral slope of the earthshine at shorter wavelengths ($\sim \lambda^{-1.25}$) could possibly be attributed to haze, soot, and/or other aerosols in the Earth atmosphere during the time of observation, which were not simulated in our models.

To test the effect of the adopted cloud particle size on the resulting spectra, we tested two additional crude exoplanet-Earth DAP models, similar to those in Figure~\ref{fig:blendcomp}, but now with type C (purple squares) or type D (green triangles) cloud particles (see Figure~\ref{fig:earthcompare}). Similar to the type B clouds used in the DAP model (solid blue line), the type C and type D cloud particles used a wavelength-independent $n_w$ of $1.335 \pm 0.00001i$ due to computational time restrictions. As discussed in Section~\ref{sec: PSDandAbs}, larger cloud particles lead to higher $P$ at most phase angles. However, while we see some  changes in the $P$ between the three DAP models, at $\alpha = 106^{\circ}$ neither of the models with larger cloud particles produced $P$ close to that of the corrected earthshine. The differences between the models and observed spectra are therefore not solely due to the cloud particle size distribution we adopted. Instead, these differences could be attributed to the fact that we assumed a single kind of liquid water clouds and did not take into account ice or mixed-phase clouds.

Water clouds on Earth show a variety of cloud particle sizes and their particle size distribution, or even phase (liquid or ice), can vary with altitude (e.g., in cumulonimbus clouds). Additionally, liquid water clouds can be covered by water ice clouds (e.g., cirrus clouds), and the global coverage of these ice clouds varies by season and latitude. The MODIS data measured on the day of the earthshine observations show a global mean coverage of $\sim$43\% for ice clouds and at least $\sim$26\% for mixed-phase clouds. Even though ice clouds have been shown to not be important for the global $P$ at some phase angles (e.g., \citealp[][]{karalidi2012rainbow} showed that they cannot mask the liquid water rainbow feature at $\alpha=40^\circ$), at $\alpha=106^\circ$ their single scattering $P$ is much larger than at $\alpha=40^\circ$ (see \citealt[][]{karalidi2012rainbow}, their Figure~3). This suggests that the effect of ice clouds on the global $P$ is stronger at the observed $\alpha$ of the earthshine and could attribute to the discrepancy between the earthshine and our models.

Finally, the observed spectrum and the two model spectra differ considerably in the appearance of the 1.27~$\mu$m $O_2$ feature. While apparent in the observations, this feature does not appear in either model. The $O_2$ feature peaks at a wavelength of $\lambda \sim 1.2685$~$\mu$m, with an increase in the polarization of $\sim 1.8\%$ above the NIR continuum in the original observations and $\sim 5.2\%$ above the NIR continuum after correcting for lunar depolarization. 
We do not reproduce this $O_2$ feature at the R of our models (R $\sim$ 250 at $\lambda = 1.27$~$\mu$m), which are slightly higher than the R of the binned earthshine (R $\sim$ 208 at $\lambda = 1.27$~$\mu$m). This presents an intriguing discrepancy between the models and observations, which we investigate further in the following section.

\section{The 1.27 \lowercase{$\mu$m} O$_2$ feature} \label{sec: o2feat}

The recent launch of the JWST and the large number of terrestrial exoplanets we will be able to characterize in the coming years means that being able to characterize biosignatures in the atmospheres of these planets is becoming increasingly important. $O_2$ is among the best biosignatures due to its photosynthetic origins and relatively higher abundance in planetary atmospheres compared to other biosignatures such as $H_2O$ \citep[][]{meadows2017}. $O_2$ produces readily detectable spectral features across the VNIR wavelengths, including the 0.69~$\mu$m $O_2$ B-band and the strong 0.76-$\mu$m $O_2$ A-band, as well as the 1.06 and 1.27~$\mu$m NIR bands, which are caused by a combination of molecular $O_2$ absorption as well as $O_2$-$O_2$ collision-induced absorption. However, $O_2$ can also be formed abiotically and could therefore constitute a false positive biosignature, making it highly important to understand the planetary environment when searching for signs of life \citep[][]{meadows2018}. For example, \citet[][]{leung2020} showed that the detection of a strong 0.69~$\mu$m $O_2$ band complimented with a weaker or undetected 1.27~$\mu$m band could indicate the presence of an ocean-loss, high-$O_2$ atmosphere, suggesting that the observed planet is no longer habitable. Nevertheless, $O_2$ remains an important biosignature as it is difficult for terrestrial planets in the HZ to maintain $O_2$-rich atmospheres without the aid of life \citep[][]{meadows2018}.

Here we analyze the 1.27~$\mu$m $O_2$ feature that was detected in the earthshine observations but was not reproduced in any of our heterogeneous Earth models (Section~\ref{sec: earthcompare}), and the implications this has for future observations. In particular, we test the effect that different resolving powers have on our models and do a parametric exploration of model parameters (T-P profiles and $O_2$ content of the atmosphere), in an attempt to match the observed $O_2$ feature in the earthshine spectrum from \citet[][]{milespaez2014}.

\begin{figure}[ht!]
    \centering
    \includegraphics[width=\linewidth]{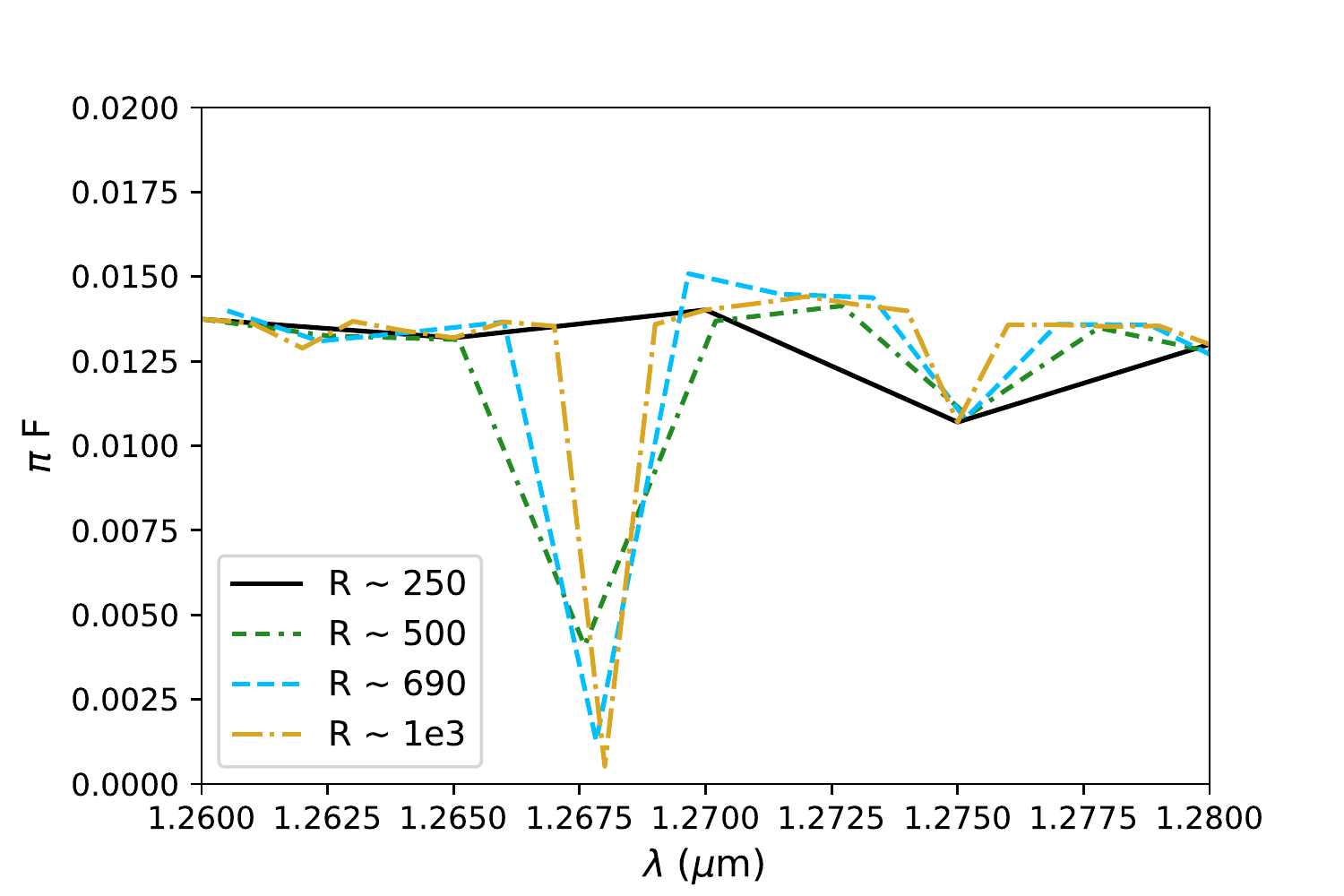}
    \centering
    \includegraphics[width=\linewidth]{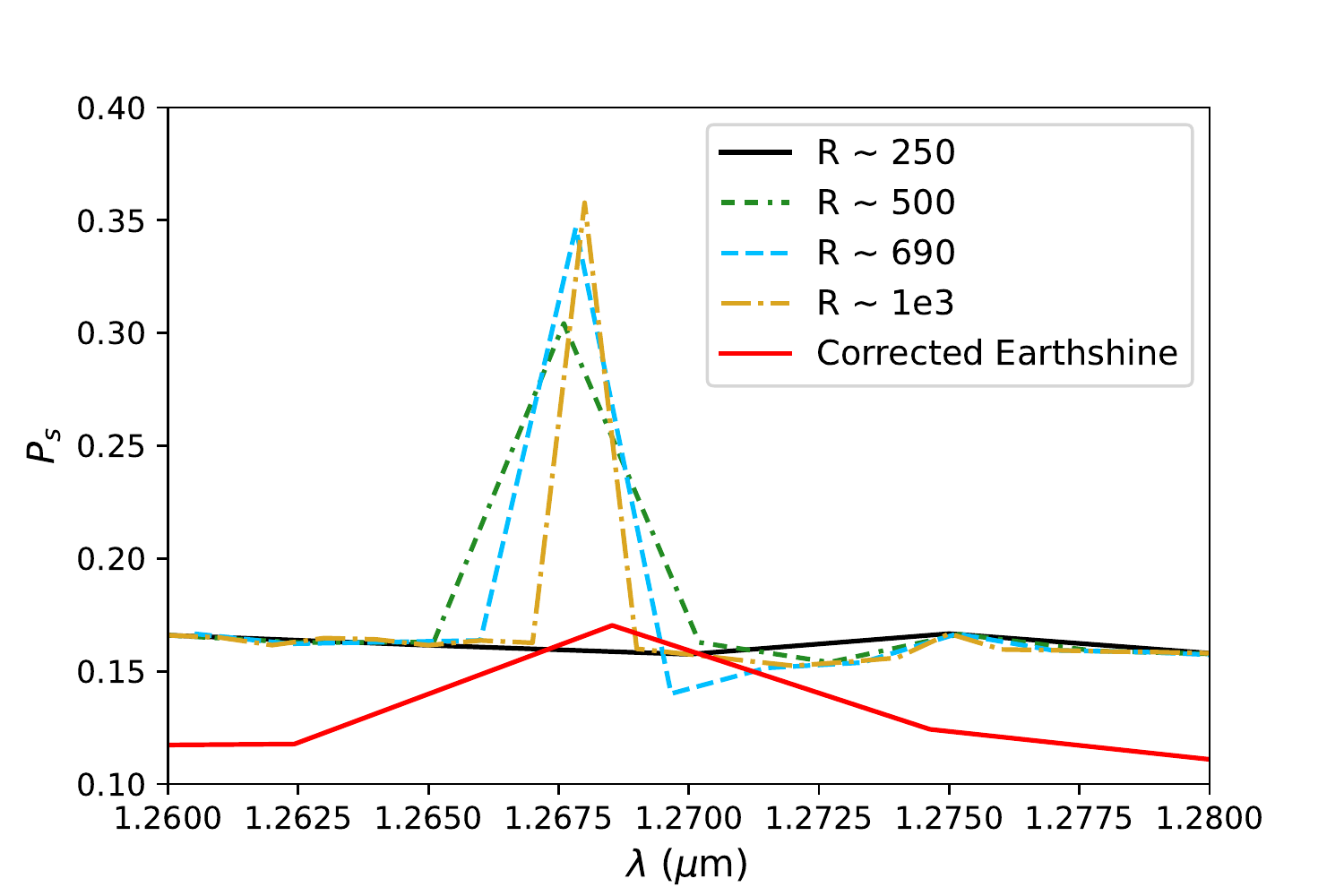}
    \caption{The wavelength dependence of model planets generated by DAP for varying R, centered around the 1.27~$\mu$m $O_2$ feature. All models are for a cloudless atmosphere above a homogeneous ocean surface, for $\alpha = 106^{\circ}$. Top panel: $\pi$F as a function of $\lambda$. Bottom panel: $P_s$ as a function of $\lambda$. The $O_2$ feature disappears in both $\pi$F and $P_s$ at R $\sim$ 250. Also included in the bottom panel is the 1.27~$\mu$m $O_2$ feature as seen in the binned earthshine observations (R $\sim$ 208, binned down from R $\sim$ 690) after correcting for lunar depolarization.}
    \label{fig:restests}
\end{figure}

\subsection{The Effect of the Model R} \label{sec: restests}

Each spectral line has a finite width and depth that are dependent on the levels of line broadening mechanisms in the planetary atmosphere \citep[][]{petty2006}. To be able to distinguish a specific feature, our instruments and models need to have larger R than the feature itself, otherwise we risk not capturing it in observed or model spectra. Here we used DAP to determine at which R we lose any signs of the 1.27~$\mu$m $O_2$ feature in our model spectra. All models were run for a planet with a clear atmosphere above an ocean surface and for a phase angle of $\alpha = 106^{\circ}$, to match the geometry of the Earth-Moon system on the night of the earthshine observations. The models cover wavelengths from 1.26~$\mu$m to 1.28~$\mu$m and have R $\sim$ 250, 500, 690, and 1000 (at $\lambda = 1.27$~$\mu$m). The R $\sim$ 250 model is at the R of the models used in previous sections, whereas the R $\sim$ 690 model represents the R of the original (unbinned) earthshine observations.

Figure~\ref{fig:restests} shows $\pi$F (top panel) and $P_s$ (bottom panel) of our model planets at the different $R$s. The $P_s$ plot also includes the 1.27~$\mu$m $O_2$ feature from the binned earthshine observations, corrected for the lunar depolarization (solid red line). Two distinct spectral lines of $O_2$ are visible in $\pi$F: one at $\lambda \sim 1.268$~$\mu$m and one at $\lambda \sim 1.275$~$\mu$m.  
As expected, decreasing the $R$ of our models results in decreased intensity and broader widths in our model lines. In $P_s$, a detectable feature is seen near $\lambda = 1.27$~$\mu$m only for the models with larger $R$. For the model with R comparable to the binned earthshine data though (R $\sim$ 250, solid black line), the feature is lost entirely due to the low number of sampled wavelengths, which causes the feature to blend with the surrounding continuum. To test that the chosen wavelength bins of the low-R model did not miss the core of the 1.27~$\mu$m $O_2$ feature, we ran additional low-R models in which we shifted the wavelength grid used to calculate the k-coefficient tables for our $O_2$ absorptions. We found that no tested wavelength grid produced a detectable 1.27~$\mu$m $O_2$ feature at R $\sim$ 250. The results here are in line with previous studies analyzing low-R \citep[e.g.,][]{brandt2014, tremblay2020} and high-R \citep[e.g.,][]{lopez2019} simulated spectra of terrestrial exoplanets in flux. Our results show that the same holds true for polarization as well.

\begin{figure}[ht!]
    \centering
    \includegraphics[width=\linewidth]{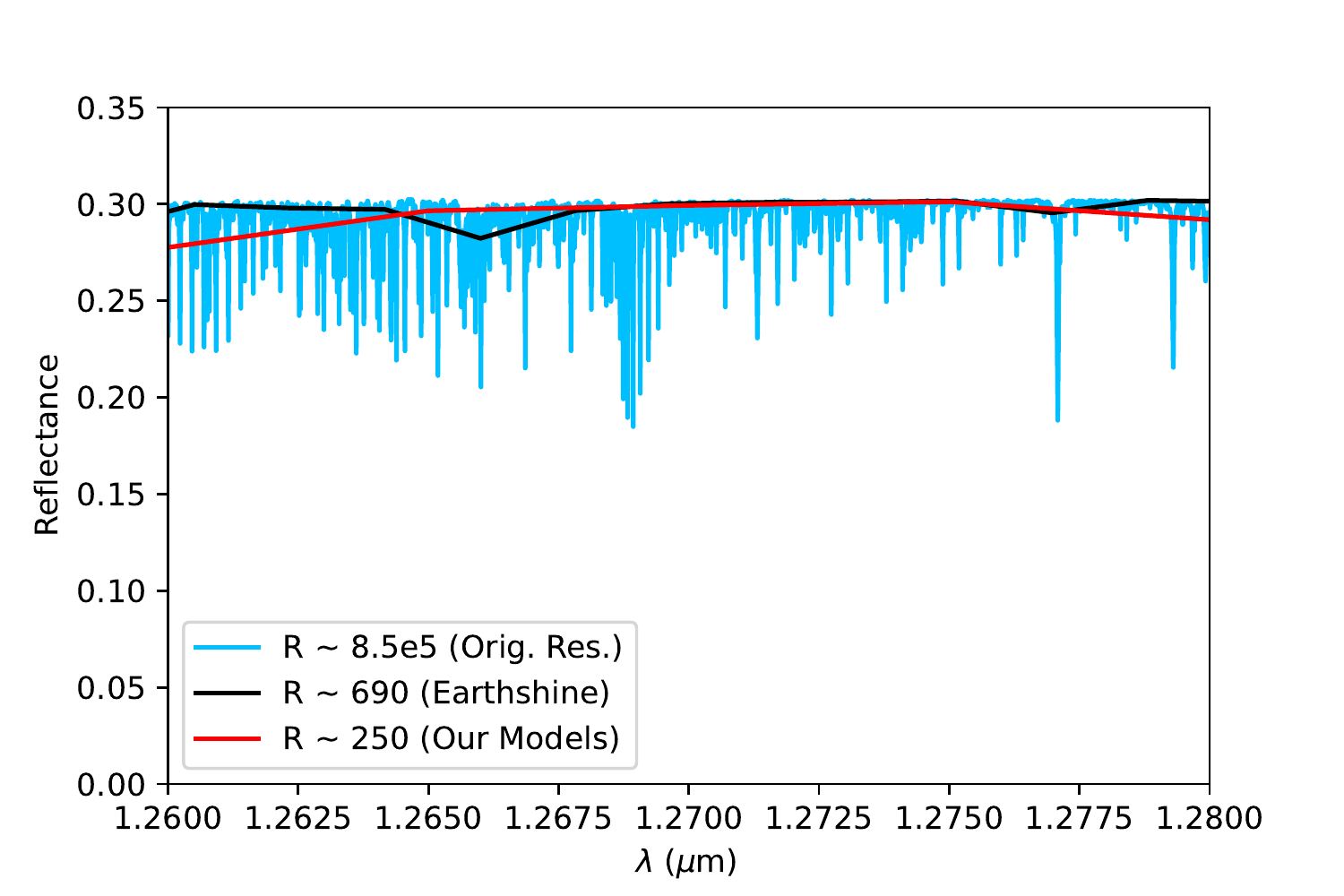}
    \caption{Simulated Earth-like atmospheric reflectance spectrum (blue line) for a planet orbiting Proxima Centauri, as generated by \citet[][]{leung2020}, centered around the 1.27~$\mu$m $O_2$ feature. The original resolution of this spectrum had R $\sim$ 850,000 at $\lambda \sim 1.27$~$\mu$m. We binned the model down to R $\sim$ 690 (black line) to match the R of the original earthshine observations from \citet[][]{milespaez2014} and R $\sim$ 250 (red line) to match the R of our models used throughout this paper. The $O_2$ spectral lines are greatly diminished in the R $\sim$ 690 spectrum and disappear in the R $\sim$ 250 spectrum.}
    \label{fig:bins}
\end{figure}

To test that our results are independent of our code and the inclusion of polarization in the radiative-transfer calculations, Figure~\ref{fig:bins} shows part of a simulated high-R reflectance spectrum of an Earth-like terrestrial planet orbiting Proxima Centauri from \citet[][]{leung2020}. The original spectrum has R $\sim$ 850,000 at $\lambda = 1.27$~$\mu$m. We binned the spectrum down to R $\sim$ 690 and R $\sim$ 250 to match the R of the unbinned earthshine spectrum as well as the R of our models from previous sections. While the 1.27~$\mu$m $O_2$ feature is detectable at R $\sim$ 850,000, in the lower R models we see only a very shallow and wide feature (R $\sim$ 690), or no feature at all (R $\sim$ 250), in agreement with our previous discussion.

\subsection{Retrieving $O_2$ from Exoplanet Atmospheres} \label{sec: o2match}

Even the highest R models that produced an observable 1.27~$\mu$m $O_2$ feature did not match that of the earthshine observations corrected for lunar depolarization (see Figure~\ref{fig:restests}). In the coming years telescopes like JWST and the Roman Space Telescope (RST) will provide us with NIR/IR spectra of multiple terrestrial exoplanets (note that JWST will only observe unpolarized light, but the RST will include a polarimeter). 
Recently, \citet[][]{fauchez2020} simulated the number of transits needed for a 5$\sigma$ detection with JWST of the $O_2$ A band (at R = 100) for a modern Earth-like cloudy atmosphere on TRAPPIST-1 e orbiting a TRAPPIST-1-like star at several distances from Earth. They found that the strong cloud opacity of their model at short wavelengths hindered the signal from this band and required upwards of 1000 transits for an accurate detection with JWST. Therefore, unless complementary ground-based observations are taken we will not have full access to the $O_2$ A- and B- bands to accurately constrain the $O_2$ content of an atmosphere. With this in mind we proceeded with scanning the parameter space of different $O_2$ abundances, surface conditions, cloud parameters, and T-P profiles to test what our retrieved planetary properties could be if we only had access to the 1.27~$\mu$m $O_2$ feature in polarization.

\subsubsection{Scanning T-P Profiles and $O_2$ Atmospheric Contents} \label{sec: tptests}

In the Earth atmosphere $O_2$ can be approximated as being evenly mixed, with a VMR of $\sim$21\%. Our models so far used this constant VMR along with a mid-latitude Earth T-P profile \citep[][]{mcclatchey1972}. However, the diversity of exoplanets observed to date suggests that different VMR profiles might exist in habitable planets. Planetary gravity $g$, T-P profiles, and VMR profiles are codependent in atmospheres. Therefore, we tested how using different $g$, T-P profiles, and VMRs in our models affect the fit of the modeled 1.27~$\mu$m $O_2$ feature with the polarized earthshine observations.

\begin{figure}[]
    \centering
    \includegraphics[width=\linewidth]{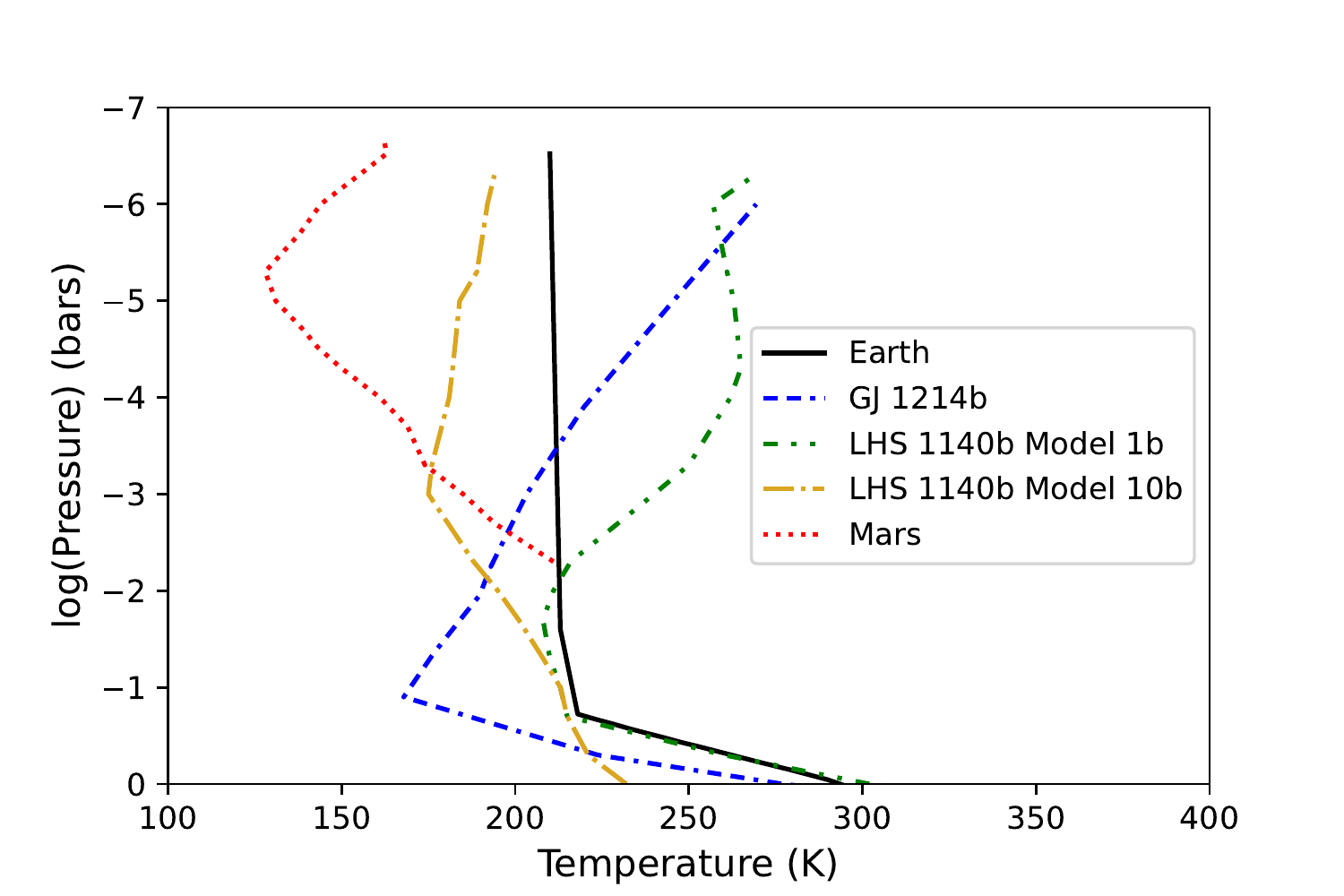}
    \caption{The T-P profiles of the five different planets modeled in this analysis, including the original T-P profile used in our exoplanet-Earth models from previous sections (solid black line).}
    \label{fig:tpprofiles}
\end{figure}

In particular, we used the observed profiles of Mars and model profiles from two of the best studied Super-Earths to date: GJ 1214b \citep[][]{charbonneau2009} and the nearby temperate Super-Earth LHS 1140b \citep[][]{dittmann2017}. Both of these exoplanets were discovered by the MEarth project survey, and due to their close distances to Earth, serve as exceptional candidates for future follow-up observations and characterizations by both ground- and space-based telescopes. For GJ 1214b we utilized the T-P profile, VMRs, and gravity of \citet[][]{morley2015} for the model with 300$\times$ solar metallicity and 0.01$\times$ the planet's incident flux, as this model was most similar to that of Earth and allowed for the condensation of water clouds in the atmosphere. 
For LHS 1140b we utilized the T-P profiles, VMRs, and gravities for two of the models of \citet[][]{wunderlich2021}: their Model 1b and their Model 10b. These models were chosen because they both have a constant CH$_4$ VMR of $1 \times 10^{-3}$, which studies have shown is expected for terrestrial planets in the HZ around mid-M dwarf stars \citep[e.g.,][]{wunderlich2019}. Additionally, Model 1b has an $H_2$-dominated atmospheric composition similar to that of Neptune, while Model 10b has a $CO_2$-dominated atmospheric composition similar to that of Mars and Venus, thereby giving us a wide range of planetary atmospheric compositions to model \citep[see Table 3 in][and references therein]{wunderlich2021}. The T-P profiles for our five scenarios are shown in Figure~\ref{fig:tpprofiles}. 
For computational efficiency we limited the number of species used in our model Super-Earth atmospheres compared to the original models. In particular, we used only $N_2$, $O_3$, $CO_2$, $CH_4$, $He$, $H_2$, $O_2$ and $H_2O$.

\begin{figure}[ht!]
    \centering
    \includegraphics[width=\linewidth]{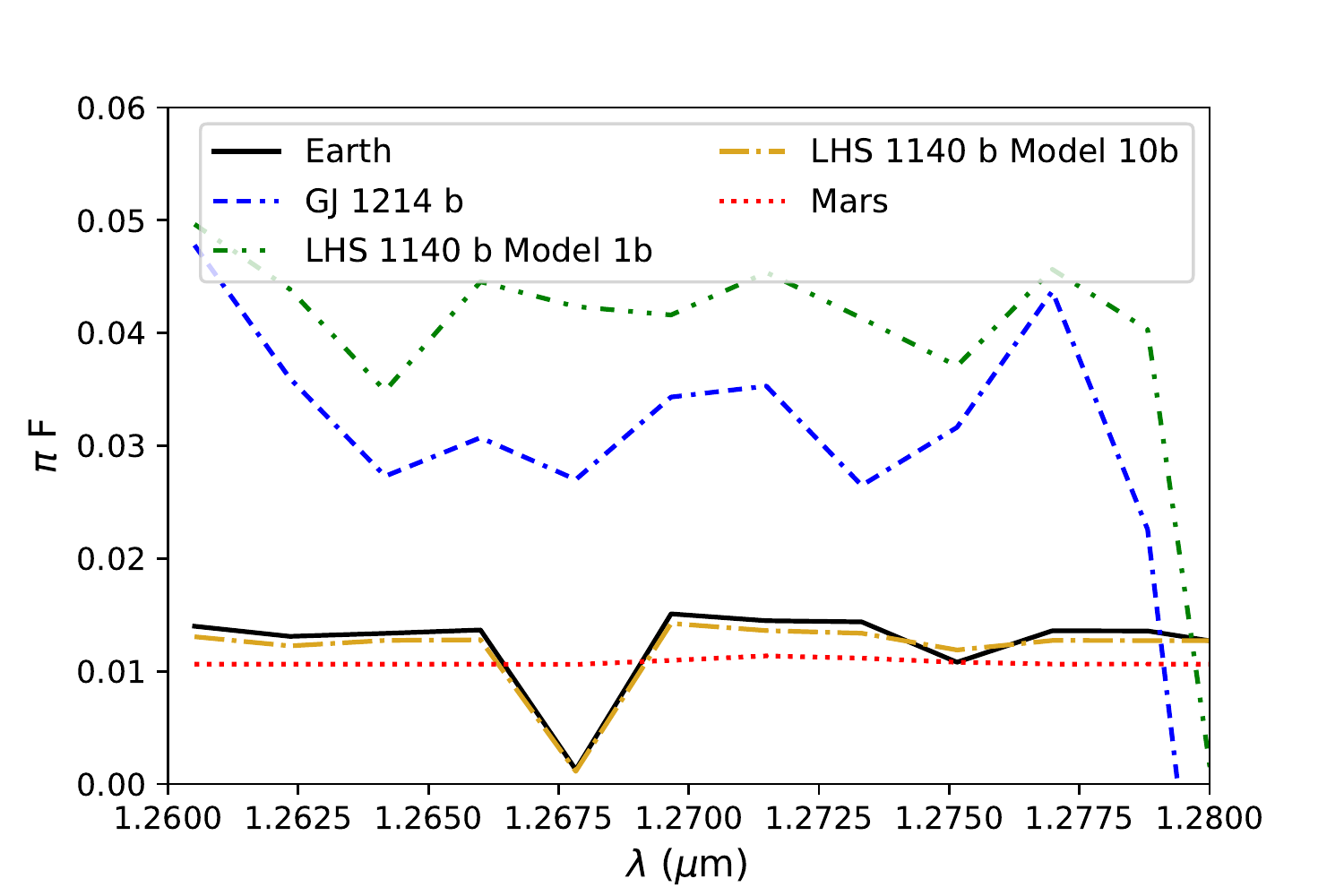}
    \centering
    \includegraphics[width=\linewidth]{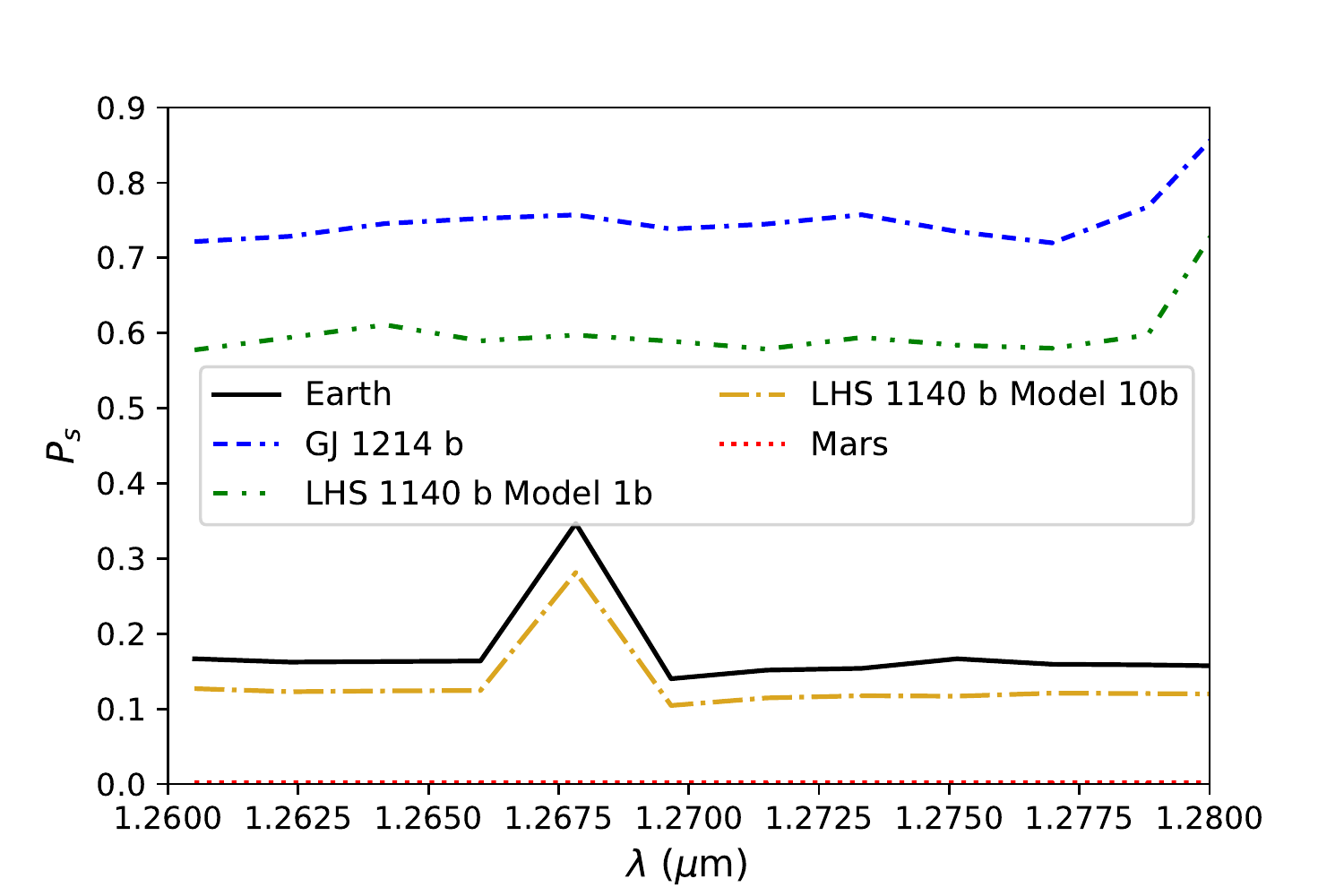}
    \caption{$\pi$F (top panel) and $P_s$ (bottom panel) of models generated by DAP for the five different planetary T-P profiles, centered around the 1.27~$\mu$m $O_2$ feature. All models are for planets with a clear atmosphere above a homogeneous ocean surface and for $\alpha = 106^{\circ}$. The models were generated at R $\sim$ 690 at $\lambda =$ 1.27~$\mu$m. Only the original Earth (solid black line) and the LHS 1140b Model 10b (dashed-dotted gold line) spectra show signs of the 1.27~$\mu$m $O_2$ feature in $P_s$.}
    \label{fig:tptests}
\end{figure}

Previous observations and analyses of these Super-Earths show that they could contain thick cloud decks, hazes, and/or potentially large levels of greenhouse effects \citep[e.g.,][]{kreidberg2014, dittmann2017}. However, clouds and hazes significantly lower the resulting polarization from a planet and require much longer computing times to model. For computational efficiency and in order to acquire higher signals that are comparable to the earthshine observations, we ran all models for cloud-free atmospheres above an ocean surface, and for $\alpha = 106^{\circ}$ to match the earthshine observations. The models cover wavelengths ranging from 1.26~$\mu$m to 1.28~$\mu$m, and were all ran at R $\sim$ 690 at $\lambda=1.27$~$\mu$m.

Figure~\ref{fig:tptests} shows $\pi$F (top panel) and $P_s$ (bottom panel) of the reflected light off our model planets. The $\pi$F of the two models with $H_2$-dominated atmospheres [GJ 1214b (blue dashed-dashed-dotted lines) and LHS 1140b Model 1b (green dashed-dotted-dotted lines)] are higher than those of the other three models. However, as these Super-Earths have atmospheres more akin to Neptune than to Earth, their models have very small $O_2$ VMRs, resulting in spectra that are featureless in both $\pi$F and $P_s$. For the Mars model (red dotted lines), the resulting $\pi$F and $P_s$ are featureless as well, which is due to the small VMRs of $O_2$ and $H_2O$ used in our Martian model. The low $P_s$ ($P_s \sim 0$) is due to the fact that the Martian atmosphere is thin and our surface was treated as Lambertian. We note that the inclusion of a realistic surface as well as dust in the atmosphere would increase $P_s$ to levels comparable to the observed Martian polarization \citep[e.g.,][]{stam2008b}. The LHS 1140b Model 10b (gold dashed-dotted lines) produces $\pi$F and $P_s$ that are both similar in shape to those of the Earth model (black solid lines), but at slightly lower values. The appearance of a strong $O_2$ feature for these two models is most likely due to the higher $O_2$ VMRs used in these models when compared to the other three: 21\% for the Earth case and 30\% for the LHS 1140b Model 10b case.

\begin{figure}[ht!]
    \centering
    \includegraphics[width=\linewidth]{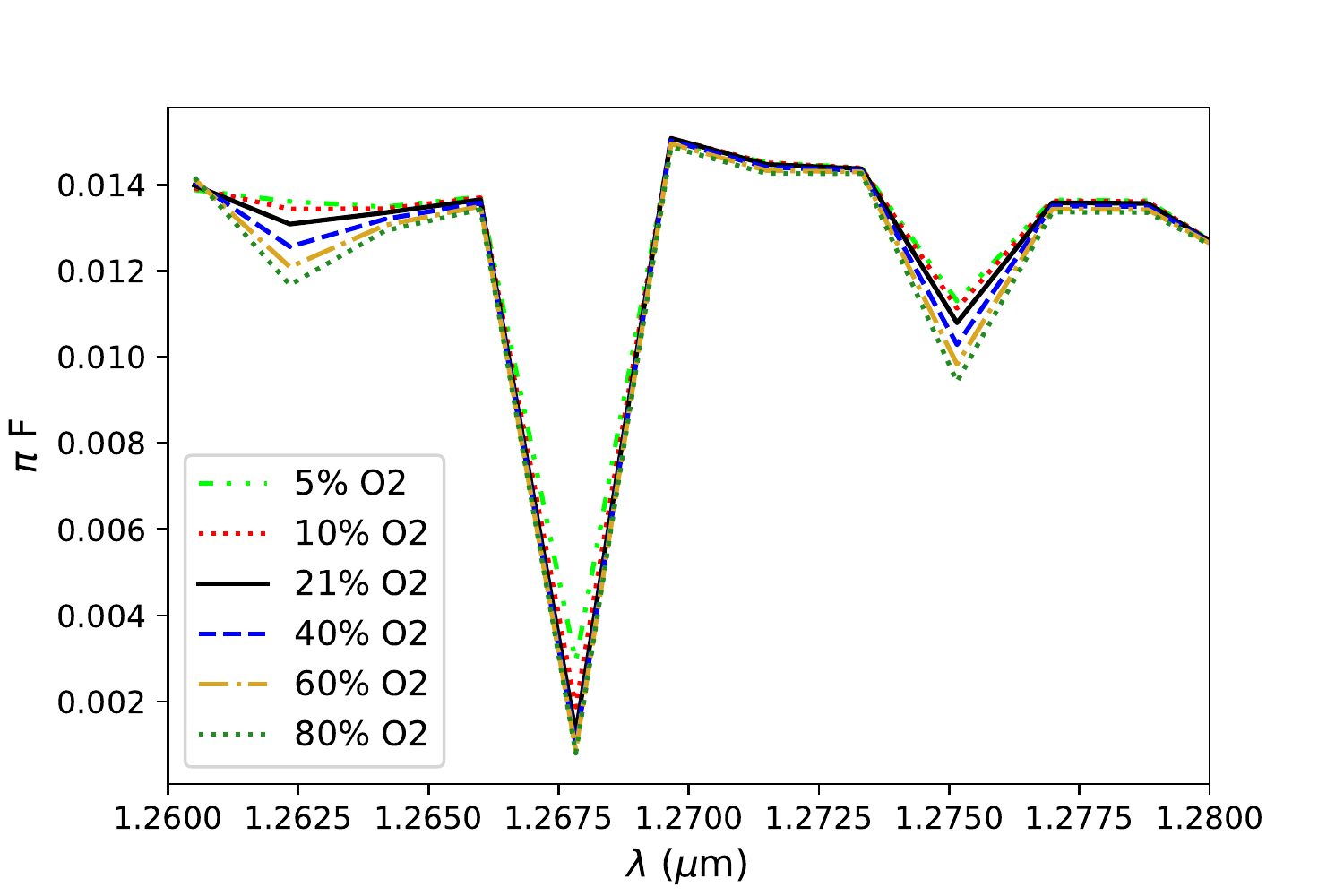}
    \centering
    \includegraphics[width=\linewidth]{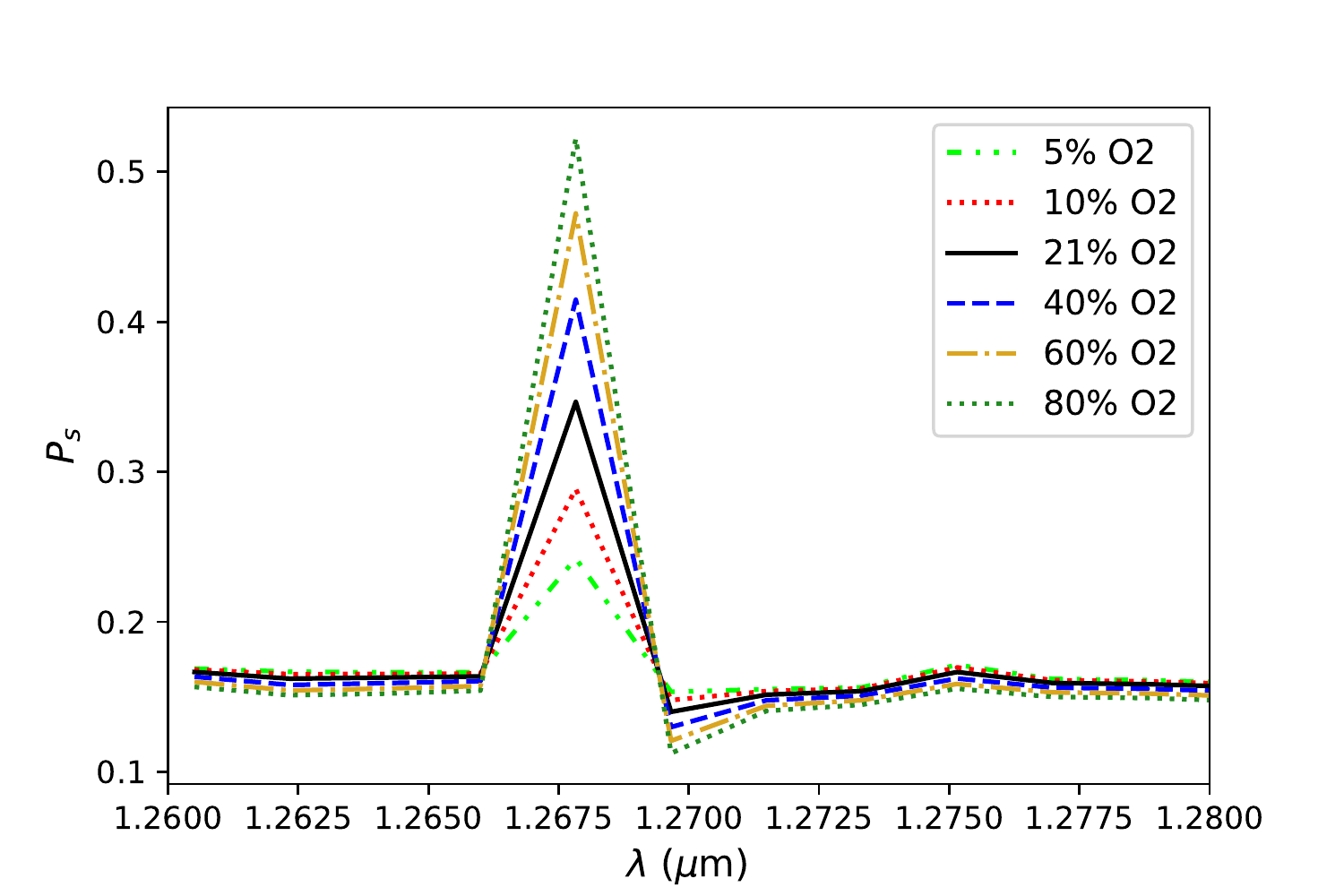}
    \caption{$\pi$F (top panel) and $P_s$ (bottom panel) of Earth models with varying levels of atmospheric $O_2$ content, centered around the 1.27~$\mu$m $O_2$ feature. All models have a clear atmosphere and a homogeneous ocean surface, and were ran at $\alpha = 106^{\circ}$. The models have R $\sim$ 690 at $\lambda =$ 1.27~$\mu$m. As expected, increasing the amount of $O_2$ in the atmosphere leads to higher $P_s$ in the 1.27~$\mu$m $O_2$ band.}
    \label{fig:O2tests}
\end{figure}

To verify this assumption, we ran additional Earth models with a range of $O_2$ VMRs from 5\% to 80\%. To ensure that the mass of the model Earth atmosphere remained the same across all models, any increase (decrease) in the $O_2$ VMR was balanced by a corresponding decrease (increase) in the $N_2$ VMR. These models have a clear atmosphere above a homogeneous ocean surface and cover wavelengths ranging from 1.26~$\mu$m to 1.28~$\mu$m, with R $\sim$ 690 at $\lambda = 1.27$ $\mu$m. Figure~\ref{fig:O2tests} shows the $\pi$F (top panel) and $P_s$ (bottom panel) of the reflected light for these models. For $VMR_{O_2}$ $\gtrsim$10\% the center of the $O_2$ feature becomes saturated in $\pi$F, so any additional $O_2$ in the atmosphere results in a negligible increase in the depth of the line. On the other hand, there is a strong correlation between $P_s$ and the $O_2$ VMR with $P_s$ increasing by $\sim$28\% (absolute $\delta P_s$) between the 5\% and the 80\% models. This confirms our assumption that higher atmospheric $O_2$ concentrations lead to a larger and more pronounced 1.27~$\mu$m $O_2$ feature in the resulting $P_s$ of our model planets.

\begin{figure}[ht!]
    \centering
    \includegraphics[width=\linewidth]{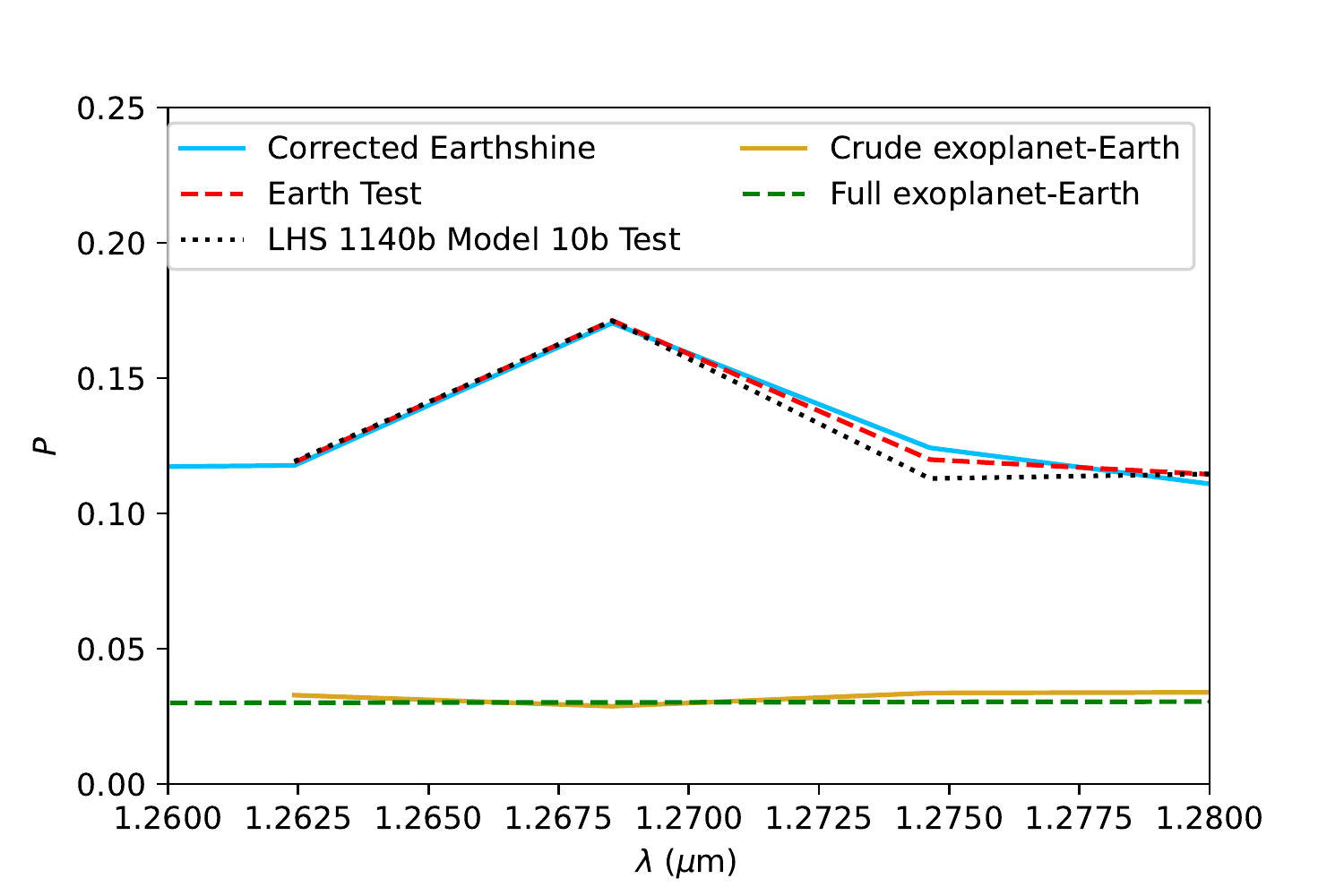}
    \caption{Comparison of the two best-fit models using the average Earth (dashed red line) and LHS 1140b Model 10b (dotted black line) T-P profiles to the earthshine observations corrected for lunar depolarization (solid blue line). The Earth model had a clear atmosphere with 10\% $O_2$ and a heterogeneous surface covered by 70\% ocean and 30\% forest. The LHS 1140b Model 10b model had a clear atmosphere with 13\% $O_2$ above a homogeneous ocean surface. The two models were both generated at R $\sim$ 690 and then binned down to R $\sim$ 208. Also included are the full-blown exoplanet-Earth DAP spectrum from Figure \ref{fig:earthcompare} (dashed green line) at R $\sim$ 250, as well as a crude exoplanet-Earth DAP spectrum (solid gold line) generated from four high-R models. Both of these exoplanet-Earth models were also binned down to R $\sim$ 208. All model spectra are for $\alpha = 106^{\circ}$.}
    \label{fig:matchfeat}
\end{figure}

\subsubsection{Matching the Observed $O_2$ Feature}

We performed a parametric scan using the different T-P profiles and varying the $O_2$ VMR between 5\% and 80\% for all model atmospheres in an attempt to match the 1.27~$\mu$m $O_2$ feature seen in the earthshine observations of \citet[][]{milespaez2014}, corrected for the lunar depolarization. Figure~\ref{fig:matchfeat} shows the two best-fit polarized spectra compared to the corrected earthshine observations (blue solid line). These best-fit spectra were generated at R $\sim$ 690 and then binned down to R $\sim$ 208 at $\lambda = 1.27$~$\mu$m. Figure~\ref{fig:matchfeat} also shows: the full DAP exoplanet-Earth spectrum from Figure~\ref{fig:earthcompare} (R $\sim$ 250 at $\lambda = 1.27$~$\mu$m; dashed green line); and a crude exoplanet-Earth spectrum (solid gold line) created from four high-resolution (R $\sim$ 1000) DAP models and generated in the same way as the crude DAP model of Figure~\ref{fig:blendcomp} (left panel).

The inclusion of any clouds in our models lowered the NIR continua, and misaligned them with respect to the corrected earthshine observations (see the exoplanet-Earth models in Figure~\ref{fig:matchfeat}). Additionally, as expected \citep[e.g.,][]{fauchez2017}, the clouds significantly reduced the intensity of the $O_2$ feature. The best-fit model ($\chi^{2}$=0.000365 vs. 0.818948 for the crude exoplanet-Earth model and 0.792116 for the full exoplanet-Earth model) that matched both the peak of the $O_2$ feature and the NIR continuum of the earthshine observations was of a planet with a clear, cloud-free atmosphere and an Earth-like T-P profile with 10\% $O_2$ (red dashed line). The model surface was heterogeneous with 70\% ocean and 30\% forest pixels. The second best-fit model ($\chi^{2}$=0.001362) was of a planet with a clear atmosphere and the T-P profile of LHS 1140b Model 10b, with 13\% $O_2$ (black dotted line). The model surface was a homogeneous ocean. Both models have lower $O_2$ VMR than their respective physically-consistent models. 
Higher $O_2$ abundances in the Earth and LHS 1140b Model 10b models caused the peak of the $O_2$ feature to rise further and mismatched the peak with respect to the corrected earthshine feature.

While we did not perform a full retrieval, rather only a limited parameter scan, these results are a cautionary tale for future observations of terrestrial exoplanets aimed at searching for biosignatures. We found that both an Earth-like planet and a Super-Earth-like planet with low $O_2$ contents could produce statistically similar polarized signals for the 1.27~$\mu$m $O_2$ feature as measured for the Earth. It will therefore be important that future studies make sure to analyze multiple bands of $O_2$ at the same time in order to truly characterize this biosignature, which could be achieved by combining ground- and space-based observations to cover the full VNIR wavelength range at high resolution.

\section{Discussion and Conclusions } \label{sec:sum}

Polarimetry of exoplanets is an underutilized tool that will help the community characterize exoplanets and break degeneracies that unpolarized light observations have. To prepare for the characterization of future polarimetric observations it is important that we benchmark our polarized radiative-transfer codes against each other and against observations of planets with known properties. Here we presented the first benchmark comparison of two polarized radiative-transfer codes, DAP and VSTAR, against each other and against polarimetric earthshine observations. Future work will build upon these comparisons by incorporating more surface and atmospheric parameters, including BPDFs, additional surface types, different cloud compositions, and additional cloud particle size distributions. 

The DAP and VSTAR models tested here showed good agreement with each other, both in the continua and in major biosignature features ($O_2$ and $H_2O$). However, a few discrepancies were noted between the models, which can mostly be attributed to the different methods of calculating the absorptions between the two codes (k-coefficient method for DAP vs. line-by-line method for VSTAR). Additionally, we noted an interesting phase dependence in the VSTAR-DAP residuals, occurring in both clear atmosphere and cloudy atmosphere model comparisons. As all of the compared models had the same surface, atmosphere, Rayleigh scattering, and (when applicable) cloud treatments, we are unsure at this time as to the cause of these discrepancies. Addressing these differences is part of ongoing work.

Comparing our models against the earthshine observations of \citet{milespaez2014}, corrected for lunar depolarization, we noted that the models underestimate the earthshine $P$, while the earthshine also has a flatter spectral slope than the models predict. This suggests that the simplifications we used affected our resulting model spectra. In particular, our assumptions of a single cloud layer per pixel and the adoption of a single cloud parameterization across the planetary disk affected the overall shape of our spectra. Our model simplifications were based on common simplifications done for modeling exoplanets. These simplifications are necessitated from the fact that an observed exoplanet will be occupying a pixel or less in our images and any rotationally asymmetric features will only be resolved through time-resolved observations. The small signal of an exoplanet-Earth would require long integration times, so that any information about small-scale variation across the disk would be lost in the observations. However, our results suggest that such simplifications can lead to large discrepancies between the data and our models. These discrepancies would have affected the retrieved properties of the exoplanet-Earth if we were using our models to retrieve the properties of the Earth using the earthshine observations.

The effect of simplifying cloud properties on exoplanet spectra has already been noted for unpolarized observations of giant planets and brown dwarfs \citep{lunamorley2021}. For terrestrial planets, \citet[][]{feng2018} showed that incorporating the full cloud 3D structure and variability is important for the proper constraint of the planetary parameters. Here we showed that the simplification of the cloud variability in an atmosphere also affects the polarimetric models of terrestrial planets. The effect of inclusion of realistic variance in cloud parameters, such as varying optical thickness with height, inclusion of ice crystals or dust particulates, variable refractive index of water ($n_w$), and varying cloud particle effective radii across the Earth's disk and with altitude will be part of future work. We note however, that increasing the complexity of models will greatly increase the computing power and time needed to run models and perform retrievals \citep[see also][]{feng2018}, so understanding which assumptions can be made to the models and in which scenarios will be important.


In Section~\ref{sec: o2match} we showed that our model R affected the fit of our models to the observed 1.27~$\mu$m $O_2$ feature. Models with R $\sim$ 250 at $\lambda=1.27$~$\mu$m, similar to the R of the binned earthshine observations, could not reproduce the $O_2$ feature at all, and only models with $R \gtrsim 500$ produced a detectable $O_2$ feature. Interestingly, even at higher R no realistic Earth-like model could reproduce the observed $O_2$ feature at 1.27~$\mu$m after correcting for lunar depolarization.
We performed a parametric scan of a range of planetary properties (T-P profiles, $O_2$ VMR) to constrain the nature of the exoplanet-Earth assuming that we knew nothing about the planet and only had access to the 1.27~$\mu$m $O_2$ feature.

We found that two models could fit the feature: an Earth-like model and a Super-Earth-like model with clear atmospheres and lower $O_2$ VMRs than the Earth. Our results highlight the importance that the synergy between ground-based Extremely Large Telescopes and space-based observatories (such as JWST and RST) will have for the full characterization of terrestrial exoplanets, even when polarimetric observations are available. Specifically, complementary observations of the $O_2$ A- and B-bands would hint on the nature of the 1.27~$\mu$m feature and aid in the confirmation that the observed $O_2$ features originate from an Earth-like planet.

While the 1.27~$\mu$m $O_2$ feature is an important biosignature and is easily detectable in Earth  spectra, the large strength of this feature in the relatively low-R earthshine spectrum of \citet[][]{milespaez2014} is intriguing. The fact that this feature could not be matched by any of our Earth-like models suggests that the depth of this observed band could be anomalous. Observations from NASA's Solar Terrestrial Relations Observatory (STEREO) and Solar and Heliospheric Observatory (SOHO) detected an Earth-directed Coronal Mass Ejection (CME) on the morning of 17 May 2013, a day and a half before the observations by \citet[][]{milespaez2014}. CMEs are known to cause geomagnetic storms, which can deplete $O_3$ levels in the upper Earth atmosphere for up to several days after the event \citep[e.g.,][]{jackman2001}. When $O_3$ is broken down by this UV radiation, it splits into molecular and atomic oxygen through photolysis. The CME event during the time of the earthshine observations could have caused rapid depletion of $O_3$, thereby increasing the amount of excited $O_2$ in the atmosphere and leading to a surplus of airglow on Earth. One of the strongest spectral signals of the Earth airglow is the 1.27~$\mu$m $O_2$ band \citep[][]{wayne1994}. This could have been picked up in the observations of the earthshine by \citet[][]{milespaez2014} and caused the large 1.27~$\mu$m $O_2$ feature seen in their spectrum. If their observed 1.27~$\mu$m $O_2$ feature turns out to be an abnormal detection related to the CME event, this would have great implications for the detection of biomarkers in polarization in the presence of an active parent star.

A vital path forward for understanding the polarimetric spectra of Earth-like planets would be to acquire more observations of the earthshine across the VNIR wavelengths. To date, the observations of \citet[][]{milespaez2014} are the only earthshine measurements to fully extend polarized observations into the NIR (\citet{takahashi2021} also studied the NIR polarimetry of the earthshine but only for broadband measurements). The observations of \citet{milespaez2014}, however, only covered one phase angle of the Earth and included a large $O_2$ feature at 1.27~$\mu$m that was unmatched by our models. While we couldn’t reproduce the observed earthshine polarization from \citet[][]{milespaez2014}, our models were able to reproduce the flux signal of the 1.27 $\mu$m O$_2$ feature from EPOXI observational data described in \citet[][]{livengood2011}. This hints further to an abnormal 1.27 $\mu$m O$_2$ strength on the day of the earthshine observations. Capturing the earthshine at more phase angles across the full VNIR wavelengths would help to build upon the results of \citet[][]{milespaez2014} and produce the full polarized phase curve of the Earth. These additional observations would greatly improve the benchmarking of theoretical models against observational data, and allow us to test whether the strength of their measured 1.27~$\mu$m $O_2$ feature is a commonality or an abnormal detection (e.g., due to increased solar activity). 
This would help pave the way for future characterizations of the biosignatures of Earth-like exoplanets.

\acknowledgments
We would like to thank Michaela Leung for providing us access to her simulated flux spectra of an Earth-like exoplanet around Proxima Centauri, as well as Dr. Caroline Morley for her atmospheric profiles of GJ 1214b, for our analyses of the 1.27~$\mu$m $O_2$ band. Additionally, we would like to thank Dr. Jeremy Bailey and Dr. David Crisp for their helpful contributions, comments, and advice that they provided for this work. K.G., T.K., and K.B. acknowledge the support of NASA Habitable Worlds grant No. 80NSSC20K1529. K.G. and T.K. acknowledge the University of Central Florida Advanced Research Computing Center high-performance computing resources made available for conducting the research reported in this paper (\url{https://arcc.ist.ucf.edu}). K.B. acknowledges the Hyak supercomputer cluster at the University of Washington used for producing the models reported in this paper. The results reported herein benefited from collaborations and/or information exchange within NASA's Nexus for Exoplanet System Science (NExSS) research coordination network sponsored by NASA's Science Mission Directorate. Finally, we thank the anonymous referees for helpful comments and suggestions, which resulted in improvements to our manuscript.

\bibliography{sample63}{}
\bibliographystyle{aasjournal}



\end{document}